\documentclass[aps,onecolumn,preprint,groupedaddress,superscriptaddress,showpacs,nofootinbib]{revtex4-1}

\usepackage{graphicx}
\usepackage{color}
\usepackage{amsmath}
\usepackage{amsfonts}
\usepackage{amssymb}
\usepackage{dcolumn}
\usepackage{hyperref}
\usepackage{enumerate}
\usepackage{bbold}
\usepackage{cancel}
\usepackage{mathtools}
\newcommand{\BEQ}{\begin{equation}}     
\newcommand{\BEA}{\begin{eqnarray}}
\newcommand{\BD}{\begin{displaymath}}
\newcommand{\EEQ}{\end{equation}}       
\newcommand{\EEA}{\end{eqnarray}}
\newcommand{\ED}{\end{displaymath}}
\newcommand{\vep}{\varepsilon}          
\newcommand{\vph}{\varphi}              
\newcommand{\D}{{\rm d}}                

\newcommand{\demi}{\frac{1}{2}}                  

\newcommand{\wit}[1]{\widetilde{#1}}         

\renewcommand{\vec}[1]{\boldsymbol{#1}} 

\begin{document}
  \title{Dynamical universality of the contact process}
\author{L. B\"{o}ttcher}
\email{lucasb@ethz.ch}
\affiliation{ETH Zurich, Wolfgang-Pauli-Strasse 27, CH-8093 Zurich, Switzerland}
\author{H. J. Herrmann}
\affiliation{ETH Zurich, Wolfgang-Pauli-Strasse 27, CH-8093 Zurich, Switzerland}  
\affiliation{
Departamento de F\'isica, Universidade Federal do Cear\'a, \\
60451-970 Fortaleza, Cear\'a, Brazil}
\author{M. Henkel}
\affiliation{
\mbox{Laboratoire de Physique et Chimie Th\'eoriques (CNRS UMR 7019), Universit\'e de Lorraine Nancy,}
      B.P. 70239, F--54506 Vand{\oe}uvre-l\`es-Nancy Cedex, France}
\affiliation{\mbox{Centro de F\'{i}sica Te\'{o}rica e Computacional, Universidade de Lisboa,} P-1749-016 Lisboa, Portugal} 
\affiliation{ETH Zurich, Wolfgang-Pauli-Strasse 27, CH-8093 Zurich, Switzerland} 
\date{\today}
\begin{abstract} 
The dynamical relaxation and scaling properties of three different variants of the contact process in two spatial dimensions are analysed. 
Dynamical contact processes capture a variety of contagious processes such as the spreading of diseases or opinions. 
The universality of both local and global two-time correlators of the particle-density and the associated linear responses are tested
through several scaling relations of the non-equilibrium exponents and the shape of the associated scaling functions. 
In addition, the dynamical scaling of two-time global correlators can be used as a tool to improve on the determination of the location of
critical points. 
\end{abstract}
\maketitle
%
%
%
\section{Introduction}
\label{sec:introduction}
%
%
%
A great variety of dynamical processes in complex systems describe the spreading of innovations \cite{coleman57,rogers2010diffusion}, opinions 
\cite{chwe99, leskovec07,boettcher_petitions,boettcher18_campaigns}, diseases \cite{satorras14, Helbing13, boettcher14, boettcher16, boettcher17_targeted}, 
damage \cite{majdandzic14, boettcher162, boettcher171}
or growing interfaces \cite{Henkel12,HH15,Henkel16,Kelling17,Kelling17b,Durang17}. 
In recent years, considerable progress has been made towards a more profound 
understanding of spreading dynamics and their corresponding phase transitions 
\cite{grassberger83, marro05, henkel08, satorras14, gleeson2013}. 
The natural relation between phase transitions in statistical physics and those characterising spreading dynamics in complex systems allows to 
apply methods of statistical physics to classify spreading models based on their universal behaviour, i.e.~the existence of the same characteristic 
asymptotic phenomena in different models and natural phenomena \cite{grassberger83, hinrichsen2000, marro05, Takeuchi2007, henkel08, Taeuber14}. 
More recently, advances in statistical physics suggested to study the so-called \emph{ageing} phenomenon describing universal features of 
non-equilibrium relaxation properties between various models \cite{henkel_ageing}. 
A characteristic relaxation behaviour has, for example, been identified for the \emph{contact process} (CP), 
a prominent example of the non-equilibrium directed percolation universality class \cite{enss2004,ramasco2004,baumann2007}, 
for stochastic Lotka-Volterra population dynamics \cite{chen16,Dobramysl17} or for gel-forming polymers \cite{Kohl16}.

%
%
%
\begin{figure}[b]
\begin{minipage}{0.98\textwidth}
\centering
\includegraphics[width=\textwidth]{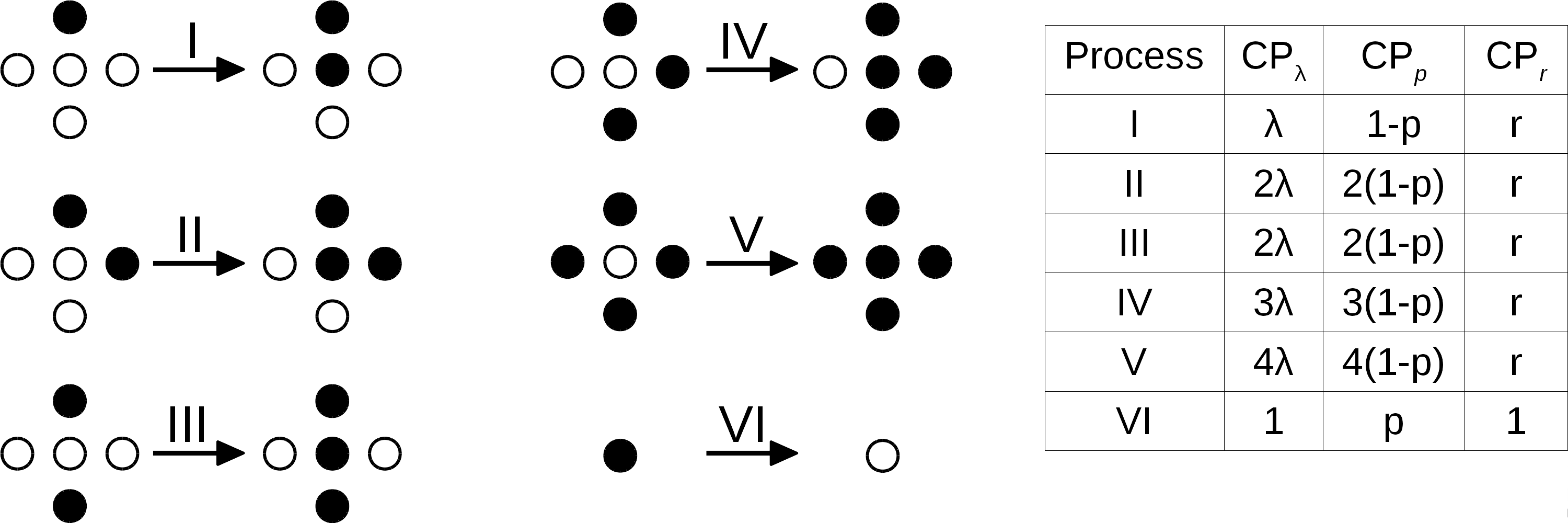}
\end{minipage}
\caption{\textbf{Update rules in $2D$ contact processes.}{ Black sites are occupied and white sites are empty. 
The rates of the microscopic processes I - VI are given in the table, for the models CP$_\lambda$, CP$_p$ and CP$_r$. 
Throughout, spatial rotation-invariance of the reactions is assumed.} 
\label{fig:cp_updates}}
\end{figure}
%
%
%
We shall study the relaxational properties of three different variants of the $2D$ CP which all have the same well-established 
stationary critical behaviour \cite{ramasco2004,marro05,boettcher162,boettcher171}. Schematically, these variants are defined in fig.~\ref{fig:cp_updates}, 
see section~\ref{sec:model} for details. 
Starting from an initial state of uncorrelated particles with a given density, 
we are interested in the relaxation of both local and global two-time correlators of the particle-density, at the critical point,  
and also in the response of the particle-density to an external addition of particles \footnote{For the contact process, the relaxation off
the stationary critical point does not lead to dynamical scaling \cite{enss2004} 
and is rather described by simple exponentials with a finite relaxation time.}.  
The associated dynamical scaling is characterised by universal exponents and scaling functions which we can test. 
Until now, such tests of the scaling far from the stationary state were carried out in the $1D$ CP universality class 
\cite{enss2004,ramasco2004,baumann2007,Henkel13a,Henkel13b}, see also ref.~\cite{Lemoult16} for recent experimental data.  
Our findings on all three realisations of the $2D$ CP agree with (and extend) earlier results on local correlators \cite{ramasco2004}
and also agree with experimental data of the turbulent liquid crystal MBBA \cite{Takeuchi2007,Takeuchi09}. 
In particular, we shall explore to what extent the existing theoretical framework with its resulting scaling relations for the exponents \cite{baumann2007}
and predictions of the form of the scaling functions \cite{Henkel06,Henkel13a,Kelling17} can be tested. 
In addition, the two-time scaling of the global correlators can be used as a tool to improve estimates for the location of critical points. 
A better understanding of CP relaxation properties might be useful as benchmark to prepare the analysis of dynamical features for 
a variety of spreading dynamics such as disease or opinion spread and thus being of great importance 
for the design of appropriate control or intervention strategies.

This paper is organised as follows. 
In section~\ref{sec:model} we describe the three variants of the CP in more detail and also introduce the necessary mathematical methods 
to analyse the scaling behaviour of the corresponding correlators and responses. All simulational results are presented in section~\ref{sec:results}. 
In section~\ref{sec:derivations} we derive the correlator and response function scaling forms. 
A summary of our results and our conclusions are given in section~\ref{sec:discussion}.
%
%
%
\section{Model and Methods}
\label{sec:model}
%
We study the relaxation characteristics of three models, defined on a square lattice, 
whose steady-state critical behaviour is identical to the one of the CP. 
Their steady-state universality has been thoroughly checked numerically
\cite{marro05,majdandzic14,boettcher162,boettcher171}; here we shall consider the non-stationary critical behaviour and its universality.  
We refer to these models as CP$_{\lambda}$, CP$_{p}$ and CP$_r$. 
Based on a master equation, their update rules are formally specified in tab.~\ref{table:cp_def} 
and are further illustrated in fig.~\ref{fig:cp_updates}. 
While CP$_{\lambda}$ is the standard definition \cite{marro05} of the contact process,  
the variant CP$_{p}$ can be computationally more efficient and has been used previously to analyse 
the relaxation behaviour \cite{ramasco2004}. The model CP$_p$ can be mapped onto CP$_\lambda$ 
by rescaling time and setting $p=(1+\lambda)^{-1}$. Despite their similarity, 
we show in section \ref{sec:relaxation} that the relaxation behaviour of CP$_\lambda$ and CP$_p$ is slightly different. 
For this reason, we consider both processes. The last model CP$_r$ 
is a special case of a general threshold dynamics whose critical steady-state is known to fall into the 
CP universality class \cite{boettcher162,boettcher171}. In contrast to 
CP$_{\lambda}$ and CP$_p$, the definition of CP$_r$ 
does not distinguish between one or more occupied neighbours, see~fig.~\ref{fig:cp_updates}.
If at least one neighbour is occupied, an empty nearest neighbour becomes occupied 
with a rate independent of the number of occupied neighbors. 
This is not the case for CP$_{\lambda}$ and CP$_p$, 
where the occupation probability increases with the number of occupied neighbours.

%
%
%
\begin{table}
\caption{Different contact process implementations on the square lattice.}
\begin{tabular}{ | p{5.5cm} | p{5.5cm} | p{5.5cm} | }
\hline
Model I: CP$_\lambda$ \cite{marro05} & Model II: CP$_p$ \cite{ramasco2004} 
& Model III: CP$_r$ \cite{majdandzic14,boettcher162,boettcher171}\\ \hline
1. Empty nearest neighbours of an occupied site become occupied at rate $\lambda$. 
& 1. No dynamics occurs on empty sites. 
& 1. Empty lattice sites, with at least one occupied neighbour, become occupied, at rate $r$. \\
2. Occupied sites become empty, ~~ at unit rate. 
& 2. Occupied sites become empty with probability $p$. With probability $1-p$, 
a new particle is created on an empty nearest-neighbour site. 
& 2. Occupied sites become empty, ~~ at unit rate. \\
3. Gillespie's algorithm is used to simulate the dynamics \cite{gillespie76,gillespie77}. 
& 3. One time-step corresponds to the inverse number of lattice sites. 
& 3. Gillespie's algorithm is used to simulate the dynamics \cite{gillespie76,gillespie77}.\\
\hline
\end{tabular}
\label{table:cp_def}
\end{table}
%
%
%

Previous simulational studies on the CP relaxation characteristics have only 
been performed for the CP$_{p}$ model, in both $1D$ and $2D$ \cite{ramasco2004}.
In $1D$, the results agree with a transfer-matrix renormalisation group (TMRG) study \cite{enss2004} 
and a one-loop $\vep$-expansion field-theoretic study \cite{baumann2007} 
and in $2D$, the Lotka-Volterra model falls into the same universality class \cite{chen16}. 
It appears therefore timely to test thoroughly the universality of the non-equilibrium critical dynamics, 
by comparing simulational data with those of other variants of the contact process such as 
CP$_{\lambda}$ and CP$_r$.
 
Non-stationary dynamics is achieved by starting from an initial state of uncorrelated particles, with average density $\langle n_i(0)\rangle=0.8$. 
Then the control parameter is set to the critical value of the stationary state and the resulting dynamics is observed. 
Previous studies \cite{enss2004,ramasco2004,chen16} have made it clear that in this setting, 
{\em physical ageing} arises, which is defined by the following properties \cite{henkel_ageing}:
\begin{enumerate}
\item non-exponential, slow relaxation,
\item breaking of time-translation invariance,
\item dynamical scaling.
\end{enumerate}
A process which satisfies only some, but not all, of these criteria may still undergo non-equilibrium dynamics, 
but does not display physical ageing, see e.g.~ref.~\cite{Esmaeili17} for a recent example. 
The existence of dynamical scaling in physical ageing implies that the underlying dynamics should exhibit universal dynamical features. 
These are measured through two-time autocorrelators and response functions.  
In order to study these quantities, we first define the average density 
\begin{equation}
\langle n(t) \rangle := \frac{1}{\cal N} \sum_i \langle n_i(t) \rangle = \langle n_i(t) \rangle,
\label{eq:density}
\end{equation}
where we sum over all local densities $n_i(t) \in \{0,1\}$ representing empty or occupied lattice sites and 
${\cal N}=L^2$ for a square lattice with linear dimension $L$. Spatial translation-invariance is assumed in eq.~(\ref{eq:density}) and throughout below. 
Next, we define the two-time {\em local} and {\em global} connected and unconnected autocorrelators, 
respectively, by \cite{henkel_ageing} \footnote{This notation is distinct from the one used in \cite{enss2004,ramasco2004}.}
\begin{subequations} \label{eq:correlateurs}
\begin{align}
C(t,s)=C_{\mathrm{local}}(t,s)                    & \coloneqq\langle n_i(t) n_i(s)\rangle-\langle n_i(t) \rangle \langle n_i(s) \rangle, 
\label{eq:localcorrelator_def1}\\
\Gamma(t,s)=\Gamma_{\mathrm{local}}(t,s)          & \coloneqq \langle n_i(t) n_i(s) \rangle, 
\label{eq:localcorrelator_def2} \\
\wit{C}(t,s)=C_{\mathrm{global}}(t,s)             & \coloneqq \langle n(t) n(s)\rangle-\langle n(t) \rangle \langle n(s) \rangle, 
\label{eq:globalcorrelator_def1}\\
\wit{\Gamma}(t,s)=\Gamma_{\mathrm{global}}(t,s)   & \coloneqq \langle n(t) n(s) \rangle. 
\label{eq:globalcorrelator_def2}
\end{align}
\end{subequations}
The first product in eqs.~\eqref{eq:localcorrelator_def1} and \eqref{eq:localcorrelator_def2} 
runs over local densities with the same indices whereas the first product in 
eqs.~\eqref{eq:globalcorrelator_def1} and \eqref{eq:globalcorrelator_def2} 
also takes cross terms into account, i.e.~terms such as $n_i(t) n_j(s)$ with $i\neq j$. 
The averaging procedure $\langle \cdot \rangle$ denotes an ensemble average over time histories.
Again, spatial translation-invariance is assumed. 
We denote the local correlators as $C$ and $\Gamma$ and the global ones as $\wit{C}$ and $\wit{\Gamma}$.

In terms of disease control, the correlators defined by eqs.~\eqref{eq:correlateurs} 
represent an important tool to quantify the prevalence of an epidemic after a certain time $t$ 
as a consequence of an earlier infection at time $s$. 
As one example, the local unconnected correlator $\Gamma(t,s)$ as defined by eq.~\ref{eq:localcorrelator_def2} 
describes the probability of a disease to be locally found at time $t$ after a local infection at time $s$. 
On the other hand, the local connected correlator $C(t,s)$ defined by eq.~\eqref{eq:localcorrelator_def1} 
is not taking into account the uncorrelated time evolutions $\langle n(t) \rangle$ and $\langle n(s) \rangle$. 
The global correlators $\wit{C}$ and $\wit{\Gamma}$ describe the disease correlations similarly to the local ones, 
however, considering the disease prevalence of the whole population. The quantities $C/\Gamma$ and $\wit{C}/\wit{\Gamma}$ 
are an indication of the degree of correlation.

In ageing systems, one expects for these autocorrelators, with
 $t,s\gg \tau_{\rm micro}$ and $t-s\gg \tau_{\rm micro}$ where $\tau_{\rm micro}$ is a microscopic reference time scale, 
the following scaling behaviour \cite{henkel_ageing}
\begin{subequations} \label{eq:correlateurs_exposants}
\begin{align}
& C(t,s)           =s^{-b}f_{C}(t/s)            \;\; , \;\; f_C(y)\sim y^{-\lambda_C/z},                           \label{eq:c_local_scaling}     \\
& \Gamma (t,s)     =s^{-b}f_{\Gamma}(t/s)       \;\; , \;\; f_\Gamma(y)\sim y^{-\lambda_\Gamma/z},                 \label{eq:gamma_local_scaling} \\
& \wit{C}(t,s)     =s^{-\wit{b}}f_{\wit{C}}(t/s)\;\; , \;\; f_{\wit{C}}(y)\sim y^{-\lambda_{\wit{C}}/z},           \label{eq:c_global_scaling}    \\
& \wit{\Gamma}(t,s)=s^{-b}f_{\wit{\Gamma}}(t/s) \;\; , \;\; f_{\wit{\Gamma}}(y)\sim y^{-\lambda_{\wit{\Gamma}}/z}, \label{eq:gamma_global_scaling} 
\end{align}
\end{subequations}
where $z$ is the dynamical exponent 
and the autocorrelation exponents $\lambda_C,\lambda_{\Gamma},\lambda_{\wit{C}},\lambda_{\wit{\Gamma}}$ 
are defined from the asymptotics for $y=t/s\gg 1$ 
of the associated scaling functions. Occasionally, we also consider the complete time-space correlator defined by
$C(t,s;\vec{r}):=\left\langle n_{\vec{r}}(t) n_{\vec{0}}(s)\right\rangle - \left\langle n(t)\right\rangle\left\langle n(s)\right\rangle 
= s^{-b} F_C\left(\frac{t}{s};\frac{|\vec{r}|}{s^{1/z}}\right)$ and its scaling behaviour. 
We shall derive and test dynamical scaling relations between these exponents in section~\ref{sec:results}. 

%
%
%
\begin{figure}
\begin{minipage}{0.49\textwidth}
\centering
\includegraphics[width=\textwidth]{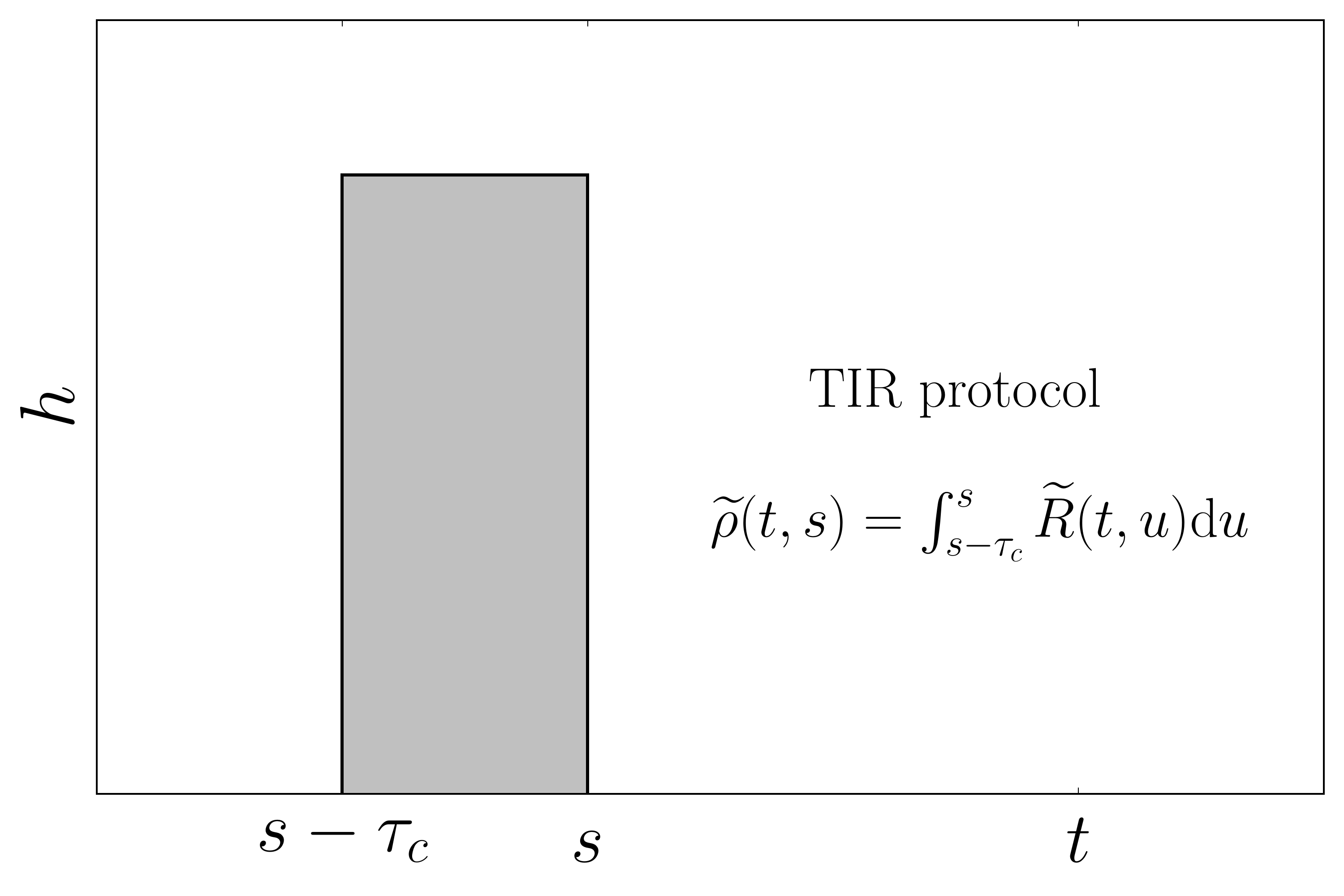}
\end{minipage}
\caption{\textbf{Field perturbation protocol.} {The time-integrated response (TIR) protocol 
is used to determine $\wit{\rho}(t,s)$. 
A field perturbation $h>0$ is applied to the dynamics in the time interval $[s-\tau_c,s]$ (grey shaded area). 
The integrated response is measured according to eq.~\eqref{eq:int_response_def}.} 
\label{fig:response_protocol}}
\end{figure}
%
%
%

Response functions can also be defined either locally or globally, according to ref.~\cite{henkel_ageing}
\begin{equation} \label{eq:response_def}
R(t,s):=\frac{\delta \langle n_i(t)\rangle}{\delta h_i(s)}\Bigr|_{h_i=0} \;\; , \;\;
\wit{R}(t,s)  := \frac{\delta \langle n(t)\rangle}{\delta h(s)}\Bigr|_{h=0}
\end{equation}
where $h_i(s)$ corresponds to the extra rate of a local creation of particles at site 
$i$ and time $s$ and $h(s) = {\cal N}^{-1}\sum_i h_i(s)$. 
While occasionally we shall consider the time-space response 
$R(t,s;\vec{r}) = \frac{\delta \langle n_{\vec{r}}(t)\rangle}{\delta h_{\vec{0}}(s)}\Bigr|_{h_i=0}
= s^{-1-a} F_R\left(\frac{t}{s};\frac{|\vec{r}|}{s^{1/z}}\right)$, 
usually we shall just consider the autoresponse $R(t,s)=R(t,s;\vec{0})$. 
By analogy with the autocorrelators, one expects a dynamical scaling behaviour of the 
response function \cite{henkel_ageing}
\begin{subequations} \label{eq:reponses_exposants}
\begin{align}
& R(t,s)=s^{-1-a}f_R(t/s) \;\; , \;\; f_R(y)\sim y^{-\lambda_R/z}, \label{eq:r_scaling} \\
& \wit{R}(t,s)=s^{-1-\wit{a}}f_{\wit{R}}(t/s) \;\; , \;\; f_{\wit{R}}(y)\sim y^{-\lambda_{\wit{R}}/z}, \label{eq:r_scaling_global}
\end{align}
\end{subequations}
along with the expected asymptotics of $ f_R(y)$ and $f_{\wit{R}}(y)$ for $y=t/s\gg 1$. 
As we shall show, the exponents of autocorrelators and autoresponses, 
defined in eqs.~\eqref{eq:correlateurs_exposants} and \eqref{eq:reponses_exposants}, 
are related by the following scaling relations
\begin{align}
b &= 2\delta \;\; , \;\; \wit{b} = b - \frac{d}{z} \;\; , \;\; \wit{a} = a - \frac{d}{z} \label{gl:scal1a}\\ 
\frac{\lambda_{\Gamma}}{z} &= \frac{\lambda_{\wit{\Gamma}}}{z} = \delta 
\;\; , \;\; \lambda_{\wit{C}}=\lambda_C-d \;\; , \;\; \lambda_{\wit{R}}=\lambda_R-d .    \label{gl:scal1b}
\end{align}
For relaxing systems with a non-equilibrium steady-state, starting from an initial non-vanishing particle density, 
the autocorrelation exponents are related to the stationary exponents as follows \cite{Oerding98,Calabrese06,Calabrese07,baumann2007}
\BEQ \label{gl:scal2}
\frac{\lambda_C}{z} = 1 +\delta + \frac{d}{z} \;\; , \;\; \frac{\lambda_{\wit{C}}}{z} = 1 +\delta.
\EEQ
Furthermore, the contact process has a specific symmetry, usually referred to as \emph{rapidity-reversal invariance}. This leads to the
further scaling relations \cite{baumann2007}
\begin{subequations} \label{gl:scal3}
\begin{align}
b &=1+a  \label{gl:scal3a} \\
\lambda_C &= \lambda_R \;\; , \;\; \lambda_{\wit{C}} = \lambda_{\wit{R}}. \label{gl:scal3b}
\end{align}
\end{subequations}
Numerical tests of these scaling relations will be presented in section~\ref{sec:results}, 
while their derivations will be given in section~\ref{sec:derivations}. 
To the best of our knowledge, previous studies solely focused on local correlators.

The response function as defined by eq.~\eqref{eq:response_def} is difficult to measure since it involves a functional derivative. 
In magnetic systems, a useful workaround is to analyse time-integrated response functions, by perturbing the system with a small external field,
which is typically chosen random in space in order to avoid introducing any bias. Models such as the contact process do not have an
easily recognised global symmetry. In such situations, it may be preferable to consider a spatially constant external field $h$, 
realised here as a particle addition rate, whose
time-dependence is illustrated in the protocol shown in fig.~\ref{fig:response_protocol} \cite{ramasco2004}. 
Then one considers a kind of damage-spreading simulation, by computing  
\begin{equation}
\wit{\rho}(t,s)=\int_{s-\tau_c}^s \!\D u\: \wit{R}(t,u) = \lim_{h\rightarrow 0}\frac{\langle n^{(B)}(t)-n^{(A)}(t;s,\tau_c)\rangle}{h}.
\label{eq:int_response_def}
\end{equation}
Herein, one compares two initially identical copies of the system which are updated by the same random numbers. 
At time $s-\tau_c$ copy A is exposed to a small field $h>0$ which is turned off again at time $s$ 
and the resulting particle-density $n^{(A)}$ is measured at time $t$. 
On the other hand, copy B remains unperturbed, giving the particle-density $n^{(B)}$. The perturbing field $h$ is applied on all sites, and the
measured particle-densities are averaged over the whole lattice, such that cross-terms between different sites appear. 
Therefore, this integrated response $\wit{\rho}(t,s)$ is indeed a {\em global} autoresponse, since
\BEQ \label{eq:rho_global}
\wit{\rho}(t,s) = \int_{s-\tau_c}^s \!\D u\: L^{-2d} \sum_{\vec{r},\vec{r}'} R(t,u;\vec{r}-\vec{r}') 
= \int_{s-\tau_c}^s \!\D u\: \wit{R}_{\vec{0}}(t,u) \simeq \tau_c \wit{R}(t,s)
\EEQ 
where $\wit{R}_{\vec{k}}(t,s)$ is the Fourier transform of $R(t,s;\vec{r})$, at momentum $\vec{k}$ and one assumes that
$\tau_c\ll s$ is small enough. 

%
%
%
\section{Simulational results}
\label{sec:results}
%
%
%
We now describe the results of the numerical simulations of the non-stationary relaxation and ageing behaviour. 
Initially, the particles are uncorrelated and for the sake of comparability with the results of earlier studies we use an average density of 
$\langle n_i(0)\rangle=0.8$ \cite{ramasco2004}. Then the control parameter
of the model (either $\lambda$, $p$, or $r$) is fixed, usually to its critical value (see eq.~\eqref{gl:point_critique}), 
and we follow the system's evolution. 
Unless stated otherwise, simulations have been performed on a square lattice with ${\cal N}=512\times 512$ 
sites and periodic boundary conditions. In tab.~\ref{tab:cp_exp} we collect our estimates for the exponents.

\subsection{Particle density}
\label{sec:relaxation}
%
%
%
\begin{figure}
\begin{minipage}{0.496\textwidth}
\centering
\includegraphics[width=\textwidth]{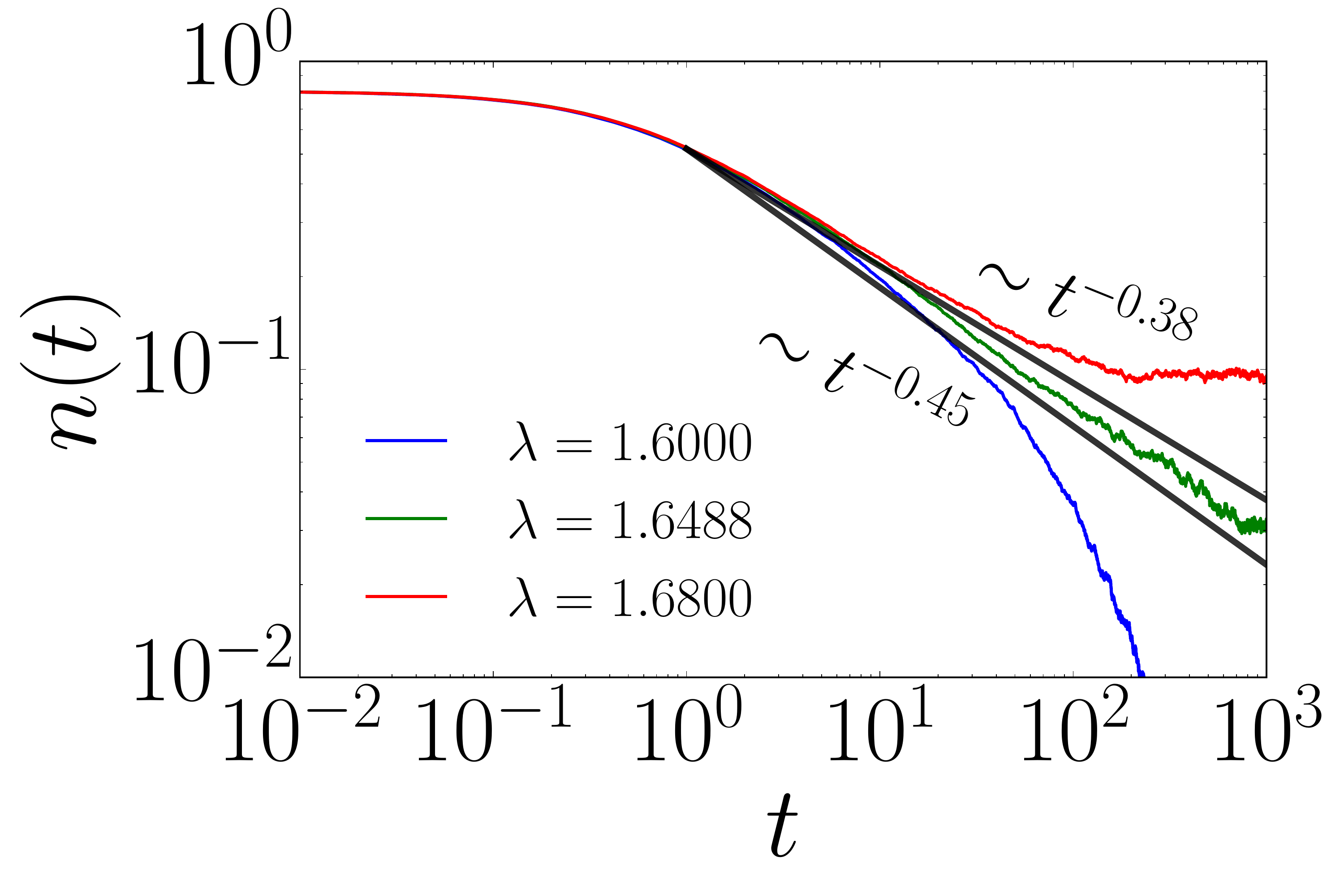}
\end{minipage}
\begin{minipage}{0.496\textwidth}
\centering
\includegraphics[width=\textwidth]{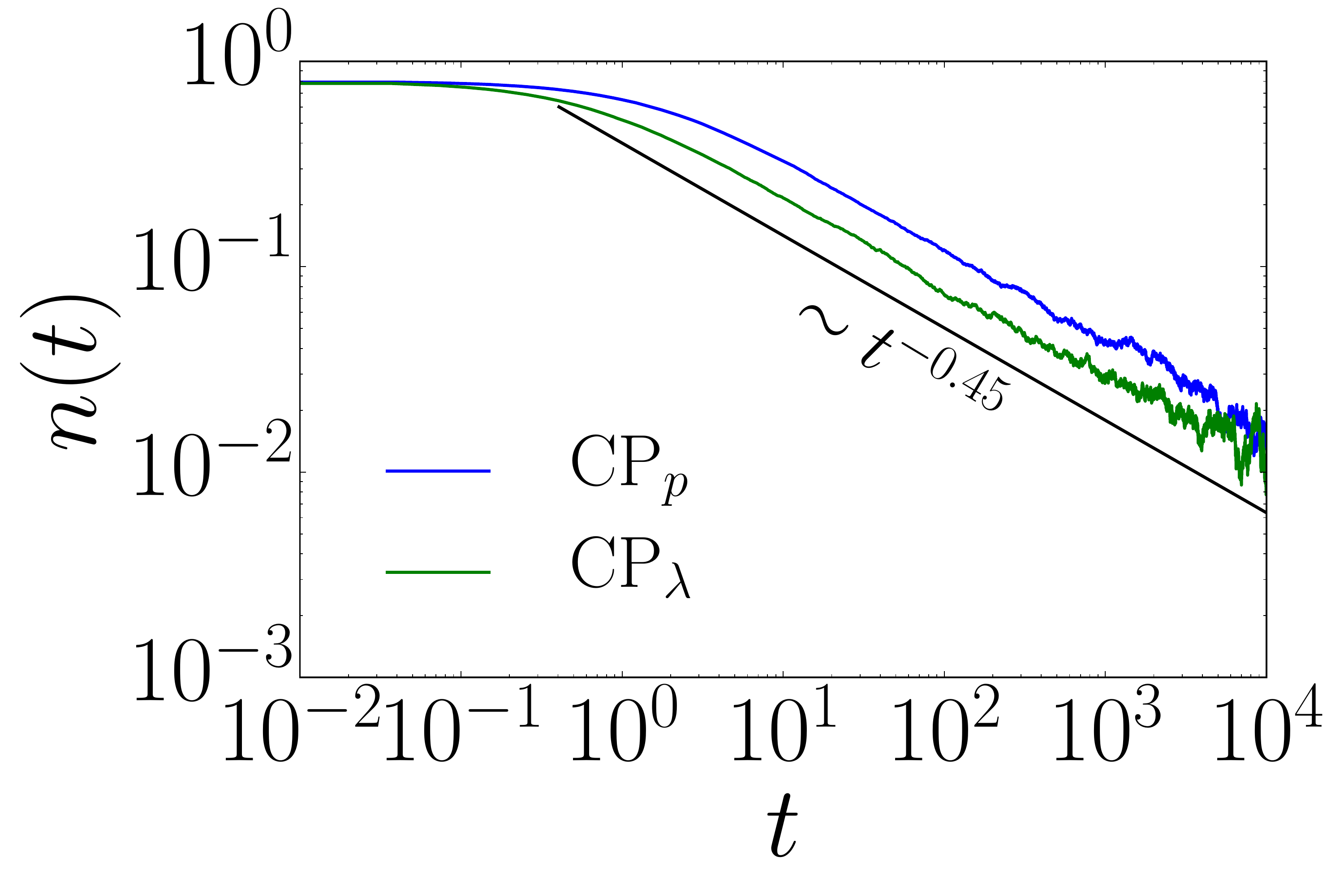}
\end{minipage}
\begin{minipage}{0.496\textwidth}
\centering
\includegraphics[width=\textwidth]{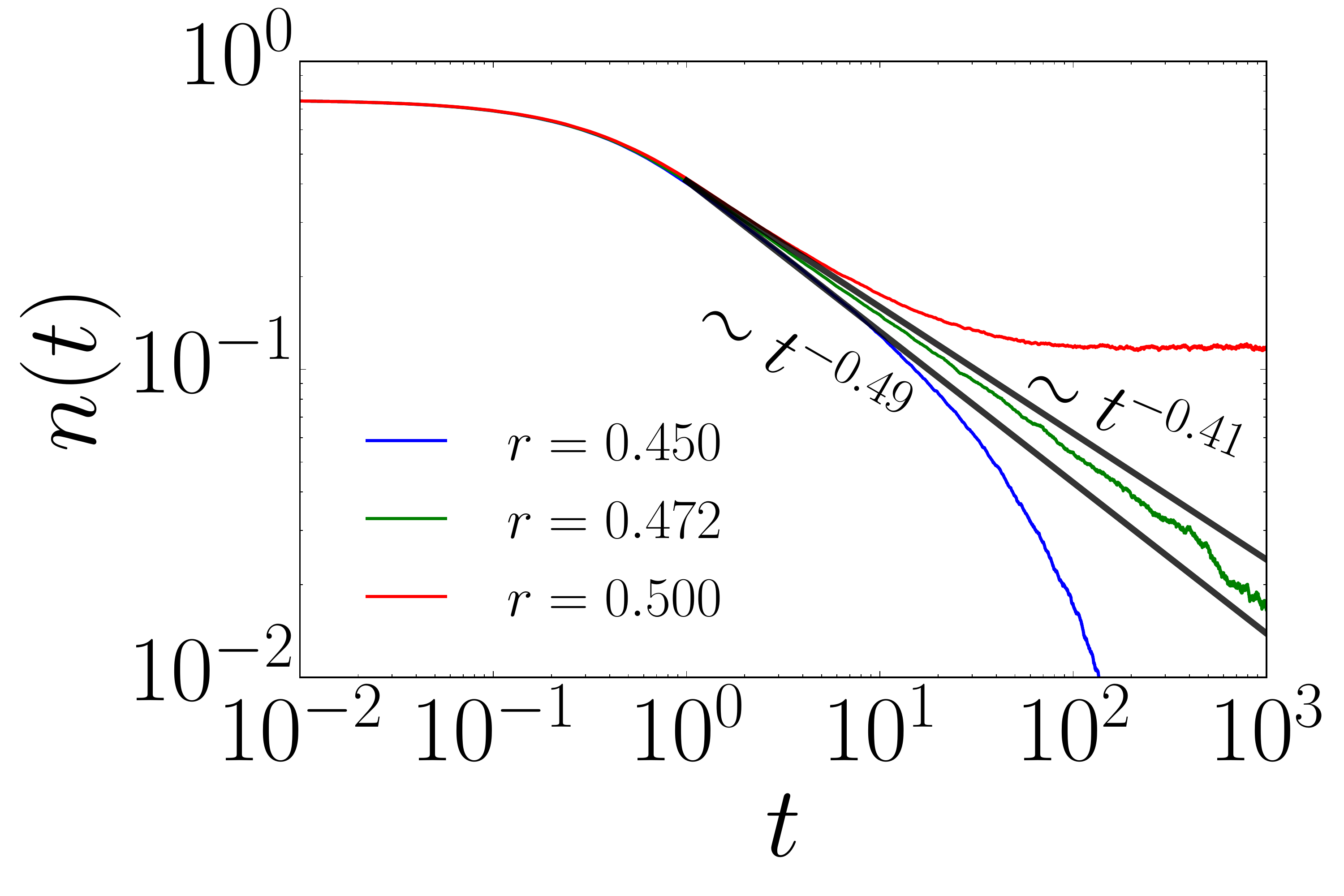}
\end{minipage}
\caption{\textbf{Relaxation of the mean particle density}. 
Upper left panel: for CP$_{\lambda}$, 
the time evolution of the density follows a power law with $n(t)\sim t^{-\delta}$ where $\delta=0.4505(10)$ \cite{voigt1997}. 
Upper right panel: time evolution of $n(t)$ for CP$_{\lambda}$ and CP$_{p}$. 
Lower panel: time evolution for CP$_r$. The critical points are given in eq.~\eqref{gl:point_critique}. 
A square lattice with ${\cal N}=1024\times 1024$ sites was used. 
The black solid lines are guides to the eye with the specified slopes.} 
\label{fig:relaxation}
\end{figure}
%
%
%
At the critical point, the power-law behaviour 
$\langle n(t) \rangle=\langle n_i(t)\rangle \simeq n_\infty t^{-\delta}$ is expected,
where $\delta=0.4505(10)$ denotes the decay critical exponent \cite{voigt1997}. 
For the three realisations of the CP under consideration, the critical points are
\BEQ \label{gl:point_critique}
\left\{ 
\begin{array}{llr}
\lambda_c = 1.6488(1)             & \mbox{\rm ~~;~ CP$_{\lambda}$~~} & \mbox{\rm \cite{moreira96},}\\
p_c=(1+\lambda_c)^{-1}=0.37753(2) & \mbox{\rm ~~;~ CP$_{p}$~~}       & \mbox{\rm \cite{ramasco2004},} \\
r_c=0.4724(1)                      & \mbox{\rm ~~;~ CP$_r$.~~}      & \mbox{\rm}
\end{array}
\right.
\EEQ
The value $r_c$ improves upon on the earlier estimate $0.47(1)$ \cite{boettcher162} by a new method which is described in the subsequent sub-section. 
In fig.~\ref{fig:relaxation} we illustrate the relaxation of all three models towards their absorbing states. 
Clearly, all three models exhibit the same power-law relaxation behaviour. 
Thus, the dynamical exponent $\delta\simeq 0.45$ is found to be the same for all three models, as expected from universality, 
and its value is in agreement with the literature. 
\subsection{Correlation functions}
\label{sec:correlators}
%
%
%
\begin{figure}[htp!]
\begin{minipage}{0.496\textwidth}
\centering
\includegraphics[width=\textwidth]{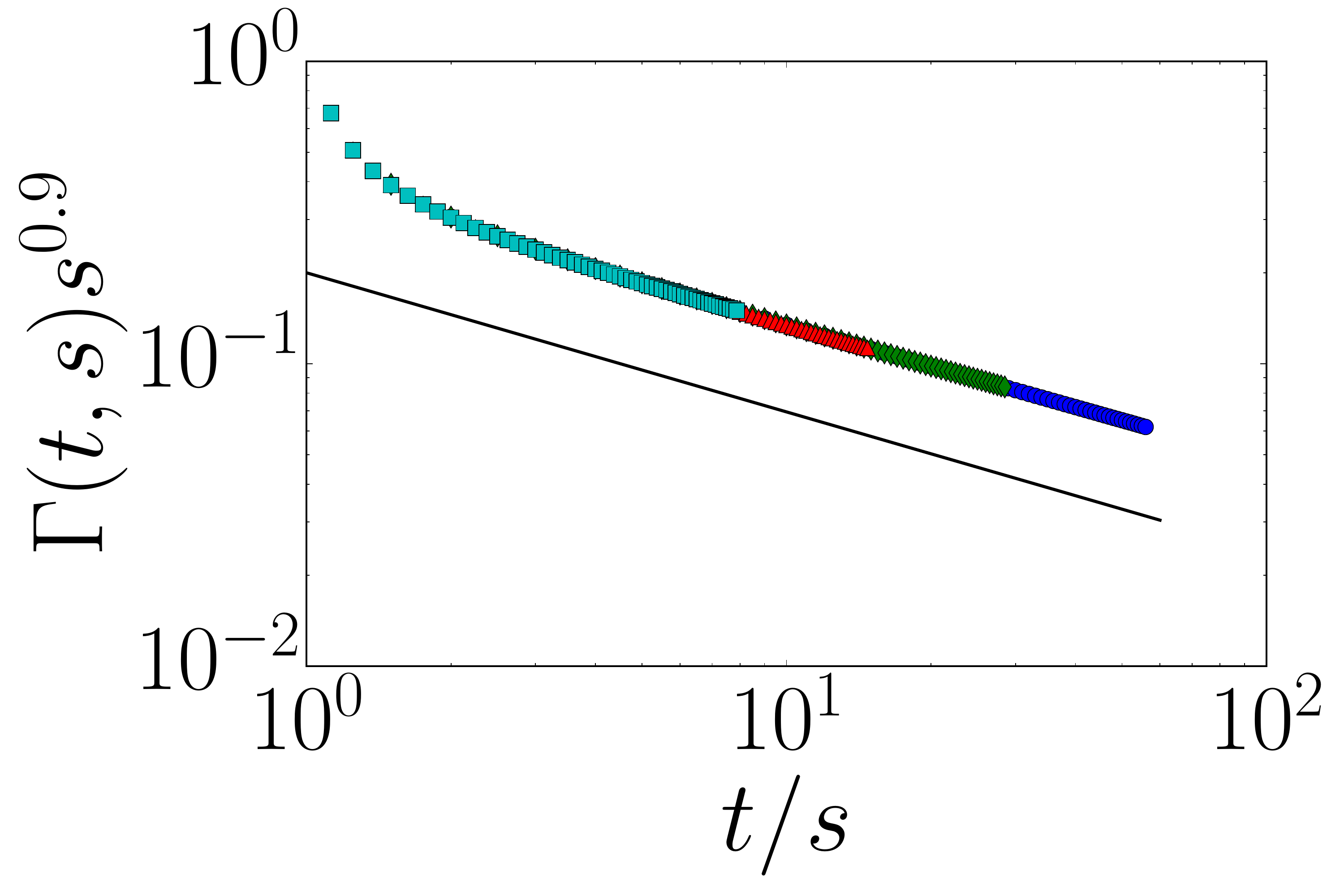}
\end{minipage}
\begin{minipage}{0.496\textwidth}
\centering
\includegraphics[width=\textwidth]{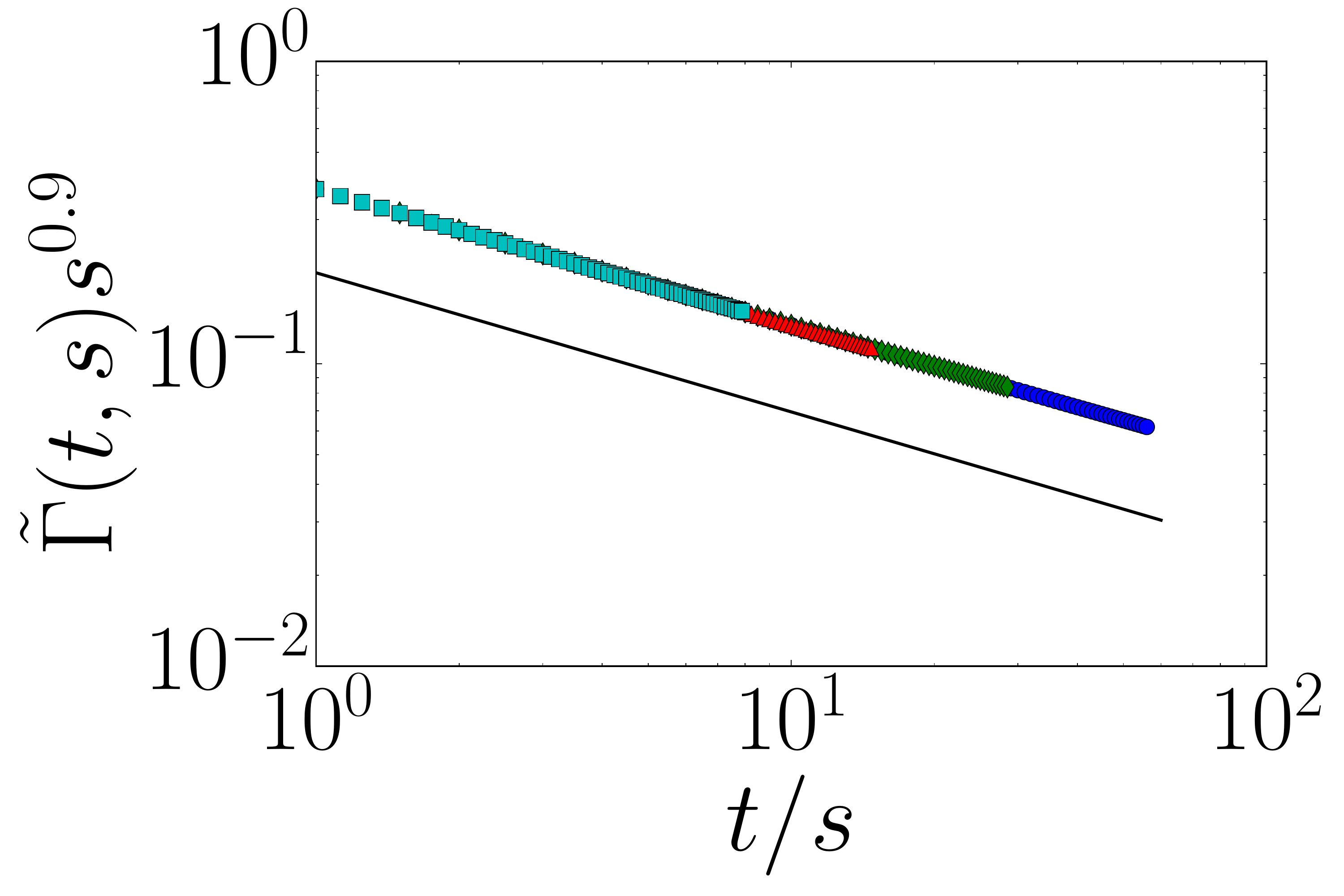}
\end{minipage}
\begin{minipage}{0.496\textwidth}
\centering
\includegraphics[width=\textwidth]{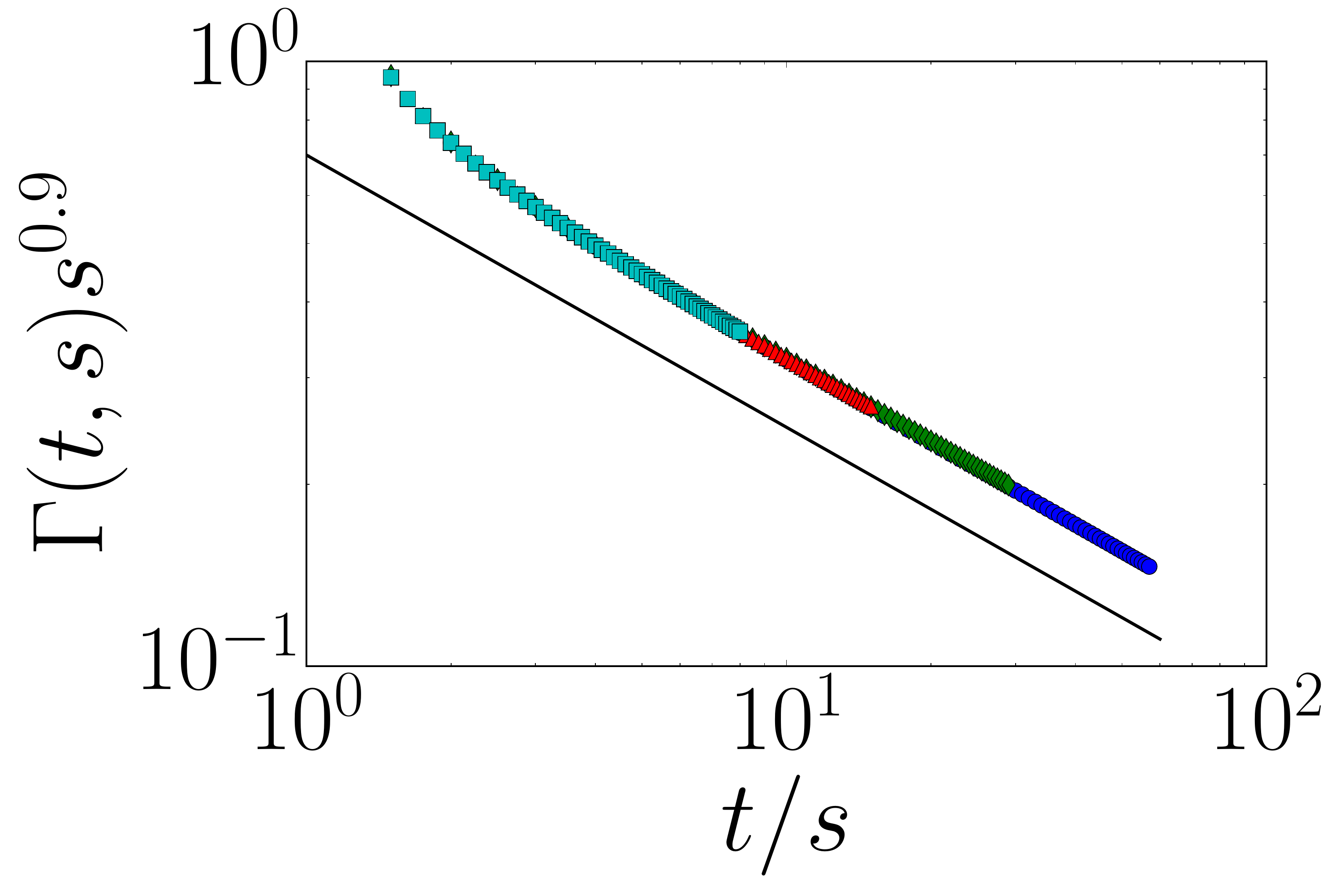}
\end{minipage}
\begin{minipage}{0.496\textwidth}
\centering
\includegraphics[width=\textwidth]{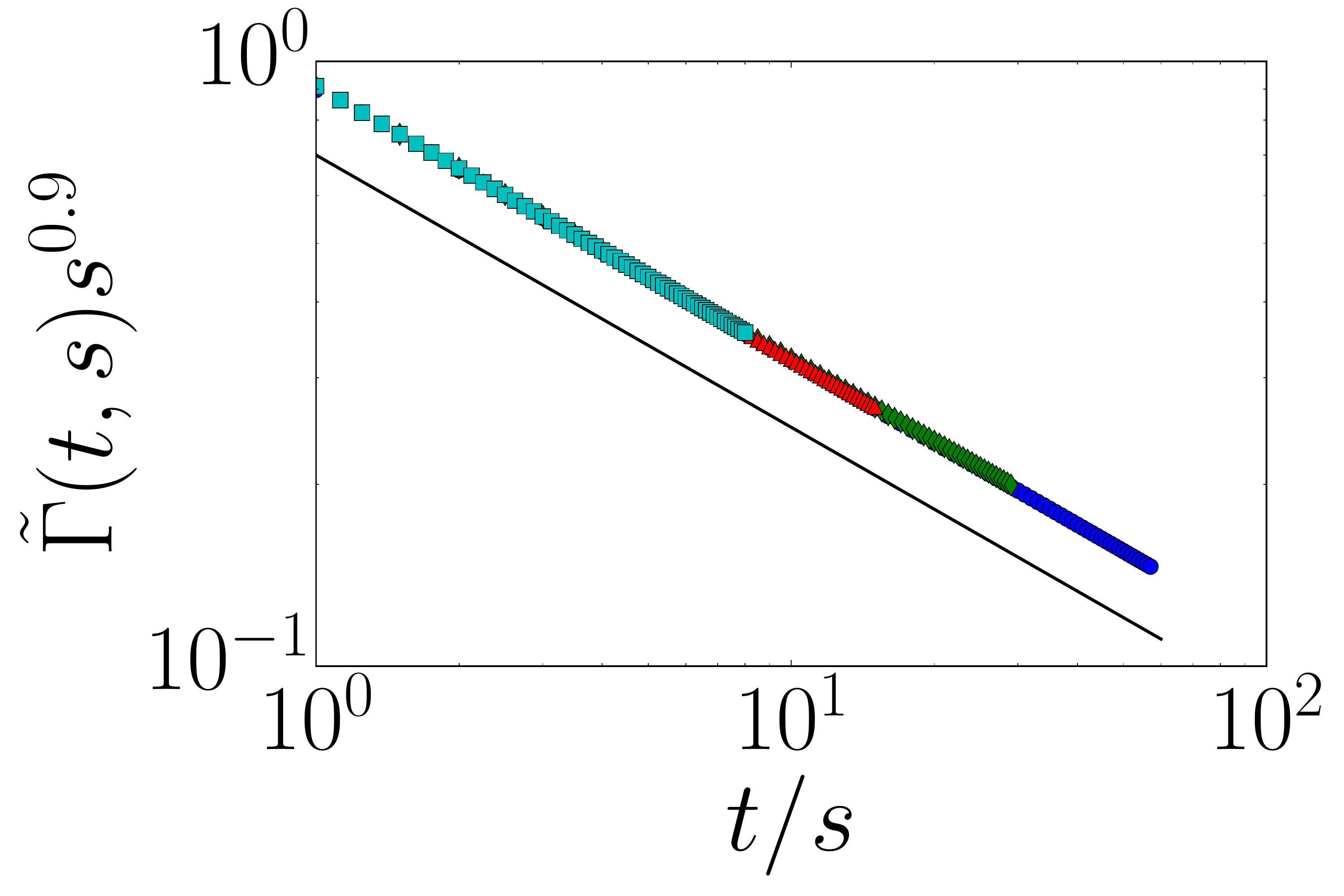}
\end{minipage}
\begin{minipage}{0.496\textwidth}
\centering
\includegraphics[width=\textwidth]{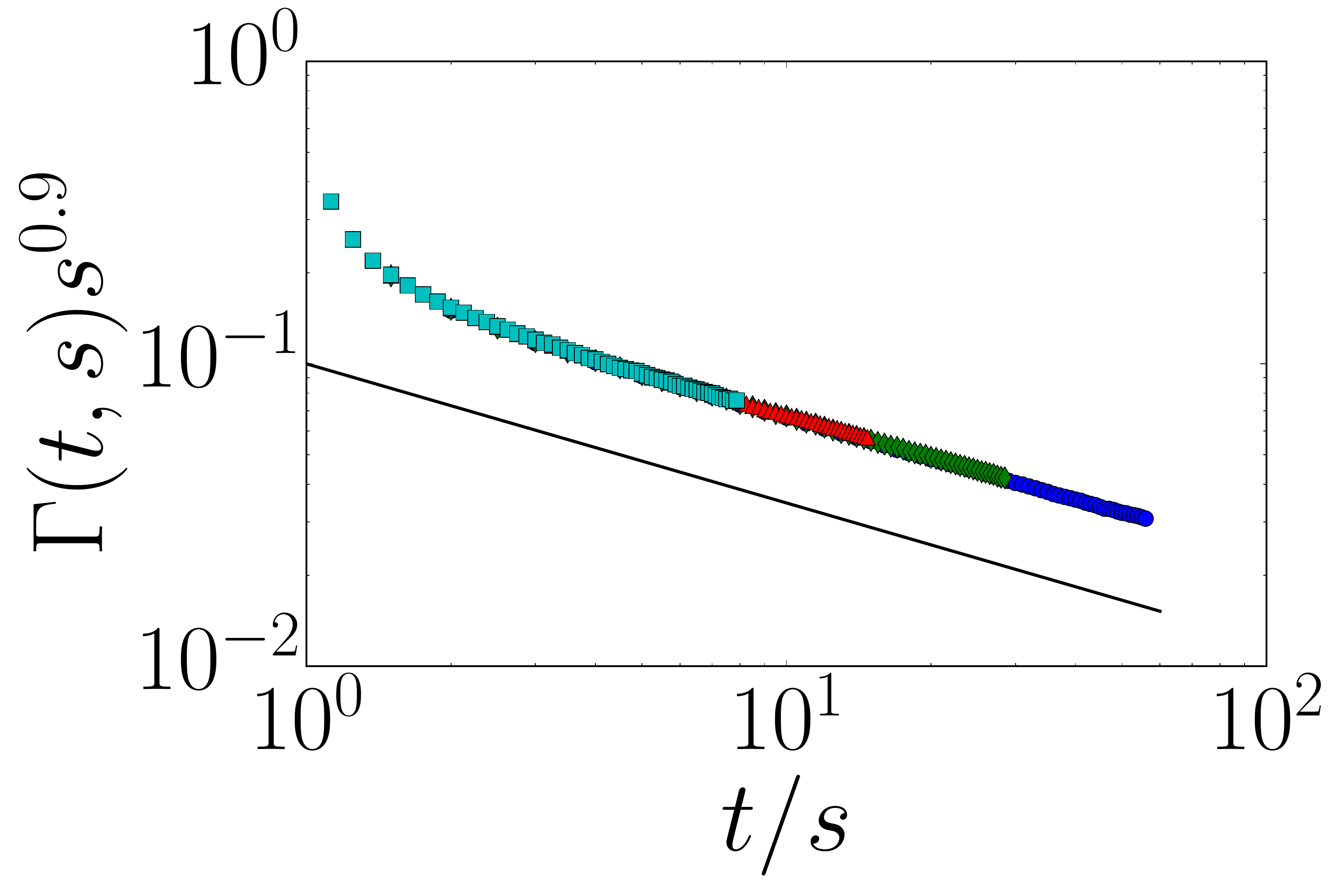}
\end{minipage}
\begin{minipage}{0.496\textwidth}
\centering
\includegraphics[width=\textwidth]{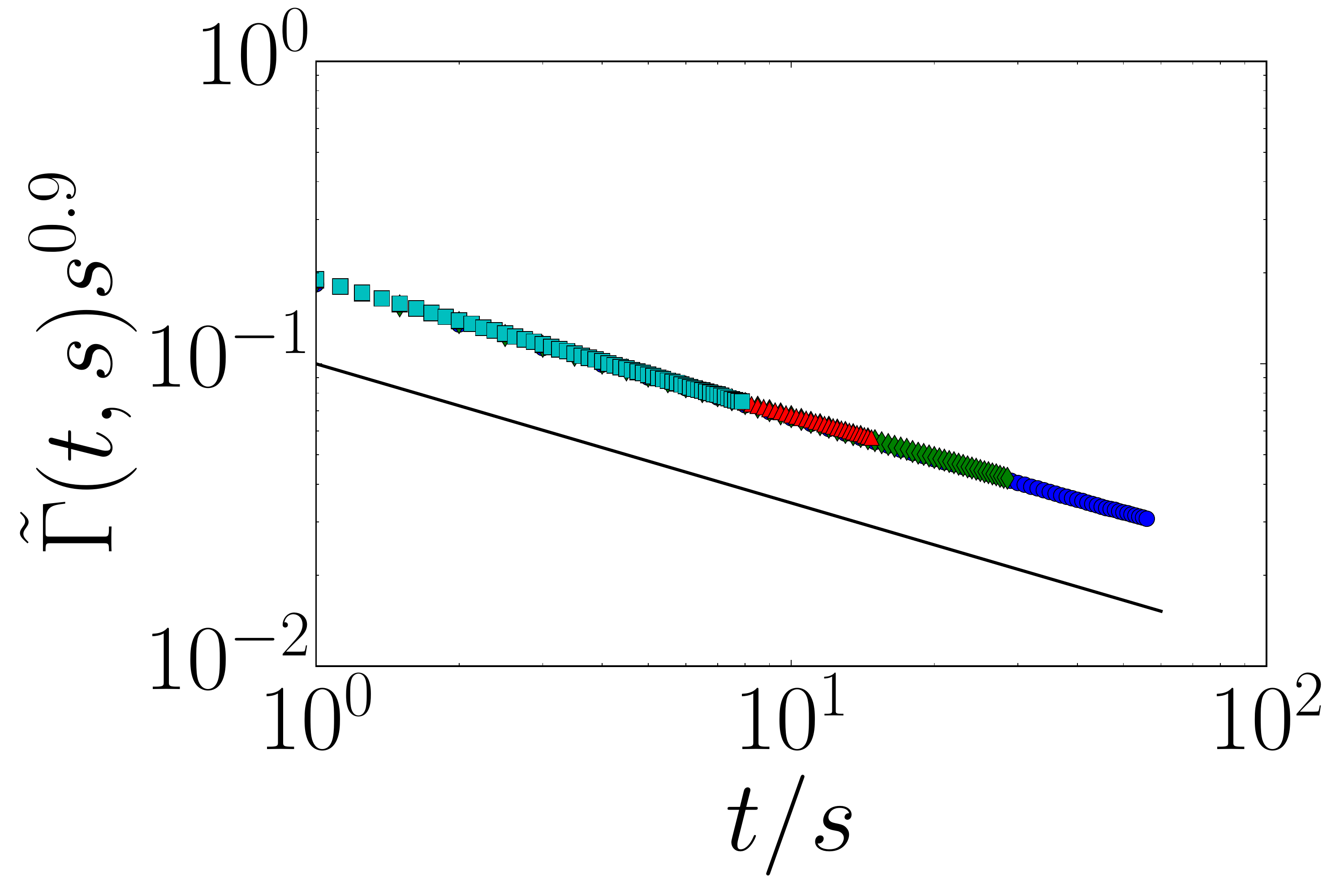}
\end{minipage}
\caption{\textbf{Unconnected local and global correlators at the critical point.} From top to bottom 
we show the data for CP$_{\lambda}$, CP$_{p}$ and CP$_r$. 
Left panels: Local unconnected correlators $\Gamma$ for different waiting times $s$. 
Right panels: Global unconnected correlators $\wit{\Gamma}$ for different waiting times $s$. 
All data were averaged over more than $10^5$ samples.
Waiting times: blue circles $s=25$, green diamonds $s=50$, red triangles $s=100$, cyan squares $s=200$. 
The black solid lines are guides to the eye with slopes $-0.45$.
} 
\label{fig:unconnected_global_local}
\end{figure}
%
%
%
\begin{figure}
\begin{minipage}{0.496\textwidth}
\centering
\includegraphics[width=\textwidth]{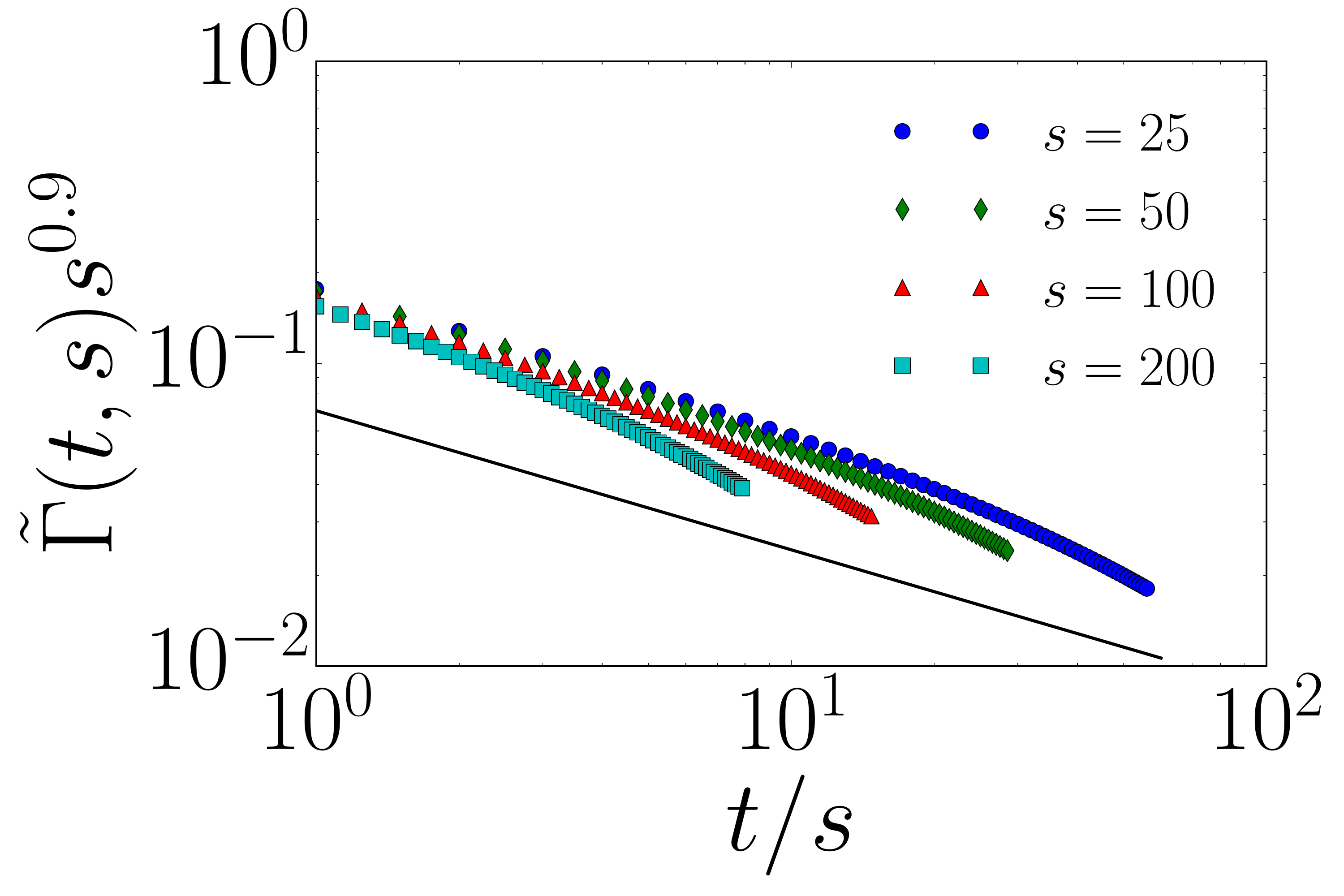}
\end{minipage}
\begin{minipage}{0.496\textwidth}
\centering
\includegraphics[width=\textwidth]{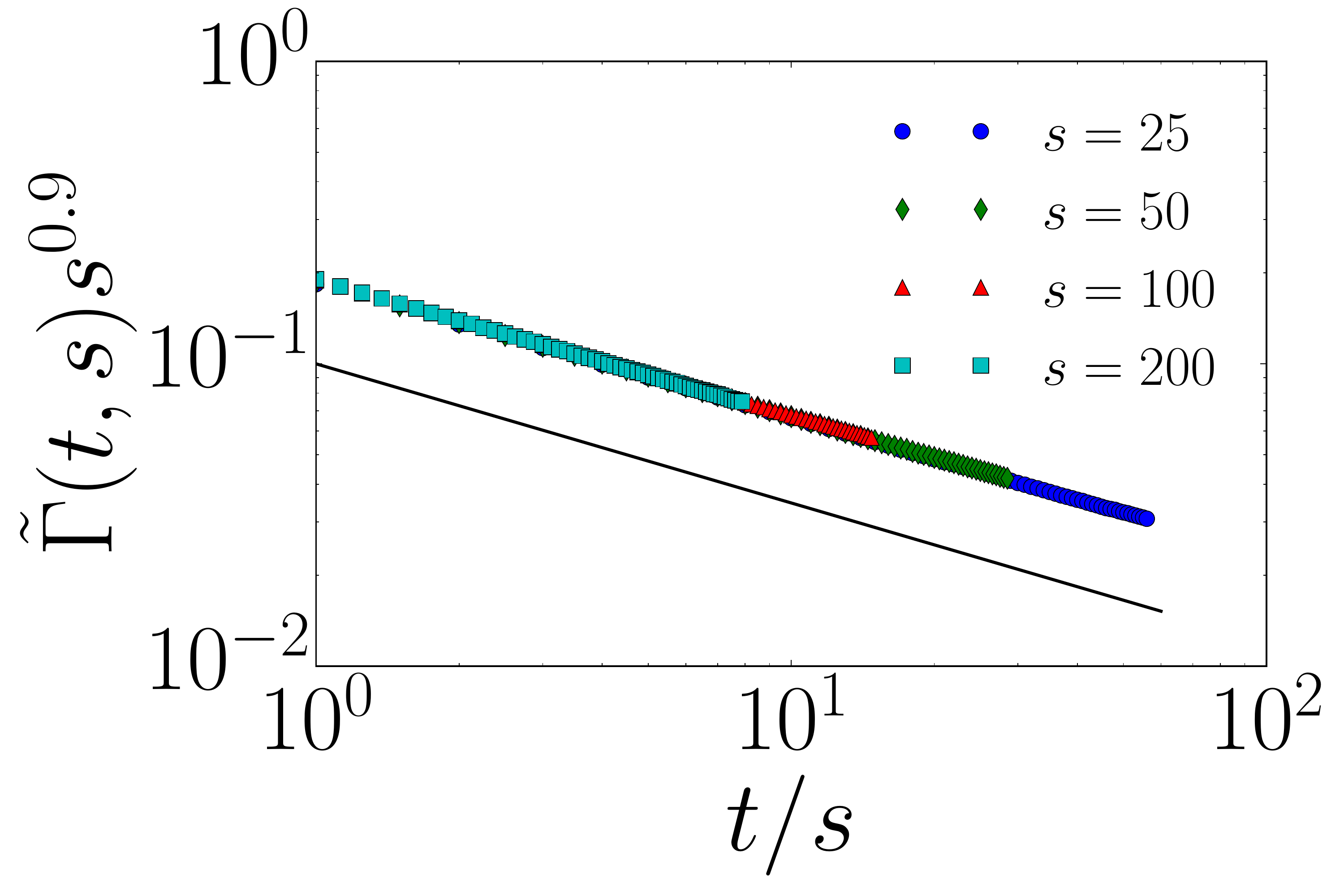}
\end{minipage}
\caption{\textbf{Unconnected global correlator dependence on deviations from the critical point}. 
For the case of CP$_r$ we show the sensitive dependence of the unconnected global correlator scaling on deviations from the critical point. 
Left panel: If one chooses $r=0.471<r_c$ deviations from dynamical scaling can be seen. 
Right panel: if one sets $r=r_c=0.4724$, a clean data-collapse results. 
All data were averaged over more than $10^5$ samples. The black solid lines are guides to the eye with slopes $-0.45$.} 
\label{fig:critical_point_m}
\end{figure}
%
%
%
We now focus on the local and global correlators as defined by eqs.~(\ref{eq:correlateurs}).  
First, in fig.~\ref{fig:unconnected_global_local} we show the unconnected correlators $\Gamma$ and $\wit{\Gamma}$, defined
by eqs.~\eqref{eq:localcorrelator_def2} and \eqref{eq:globalcorrelator_def2}, 
for the three different variants of the contact process CP$_\lambda$, CP$_p$ and CP$_r$. 
Clearly, for all three realisations the expected scaling behaviour \eqref{eq:gamma_local_scaling} and \eqref{eq:gamma_global_scaling} is seen, 
where the value of the exponent $b$ is consistent with the expectation $b=2\delta=0.901(2)$ \cite{ramasco2004}. 
For the local correlator, this is readily understood by re-writing the local unconnected correlator in terms of the connected one
\begin{equation}
\Gamma(t,s)= \langle n(t) \rangle \langle n(s) \rangle+C(t,s) 
\simeq  s^{-2 \delta} \left[n_\infty^2 \left( \frac{t}{s}\right)^{-\delta}+f_\infty \left(\frac{t}{s}\right)^{-\lambda_C/z}\right],
\label{eq:local_scaling}
\end{equation}
where we used the late-time behaviour of the density $\langle n(t)\rangle \simeq n_\infty t^{-\delta}$ and the scaling form
(\ref{eq:c_local_scaling}) for the autocorrelator. Because of the known value 
$\delta=0.4505(10)$ \cite{voigt1997} and the estimates $\lambda_C/z=2.8(3)$ \cite{ramasco2004} or 
$\lambda_C/z=2.58(2)$ \cite{baumann2007}, we have $\delta<\lambda_C/z$ such that the first term 
in eq.~\eqref{eq:local_scaling} dominates for large values of $t/s$. 
The slope of the plot is in agreement with the value $\lambda_{\Gamma}/z=\delta$, expected from (\ref{gl:scal1b}). 
An analogous, but slightly more involved argument applies to the global correlator 
$\wit{\Gamma}$ and will be presented in section~\ref{sec:deriv_correlators} and confirms the measured value of 
$b$ and of $\lambda_{\wit{\Gamma}}/z=\delta\approx 0.45$. 

An application of the scaling of unconnected correlators concerns the refinement of estimates for the critical point. 
In fig.~\ref{fig:critical_point_m} we consider the scaling of $\wit{\Gamma}$ in the CP$_r$ model at the previous estimate of
$r_c\simeq 0.471$ \cite{boettcher162}. The absence of a clear data collapse for $r$ even slightly off the precise value of $r_c$ 
is a result of the higher sensitivity of $\wit{\Gamma}$ 
with respect to small deviations from $r_c$ as compared to the particle density $n(t)$ since much larger time scales are explored. 
Indeed, at the new estimate $r_c=0.4724(1)$ a perfect data collapse is observed as shown in fig.~\ref{fig:critical_point_m}.
%
%
%
\begin{figure}[!htp]
\begin{minipage}{0.496\textwidth}
\centering
\includegraphics[width=\textwidth]{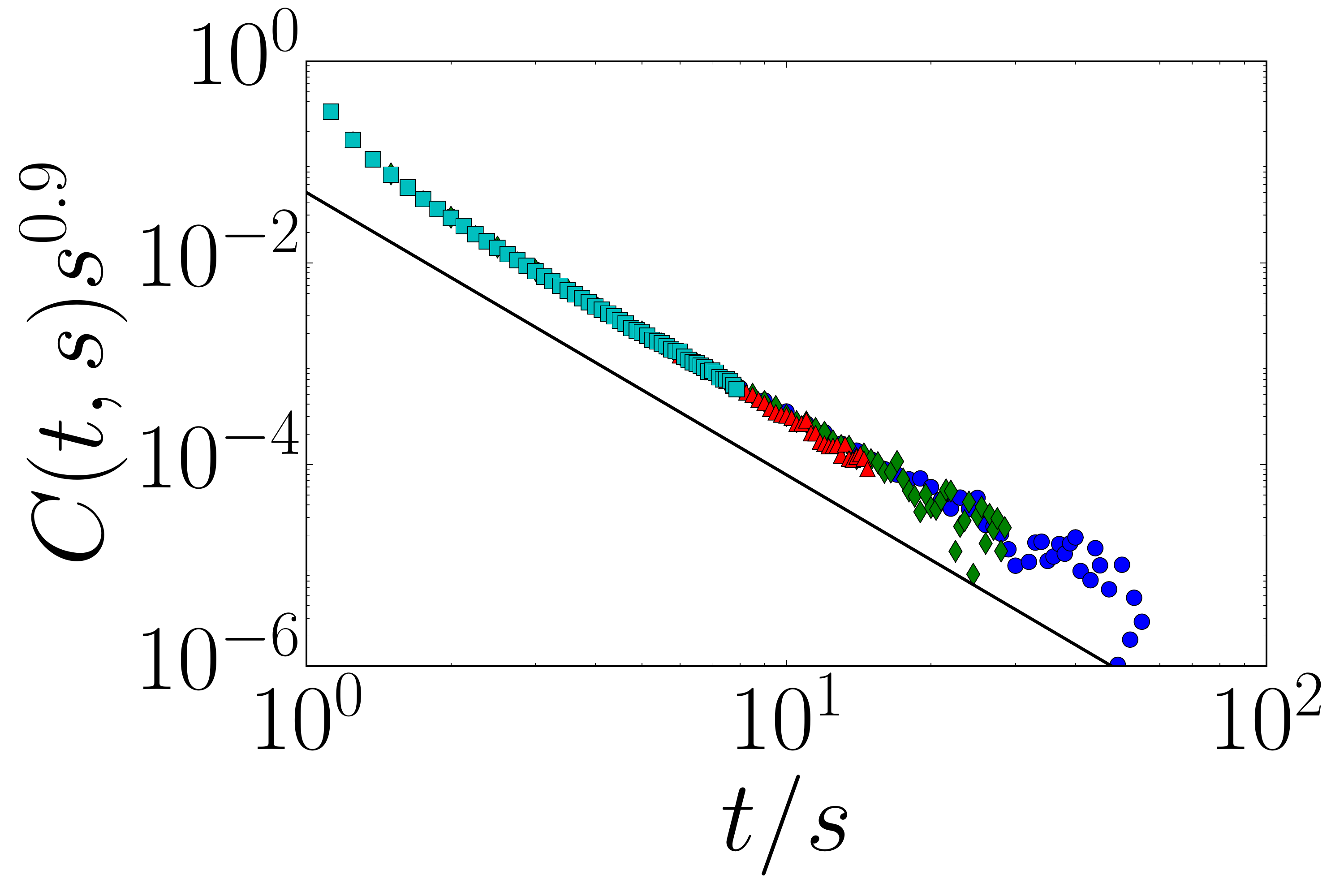}
\end{minipage}
\begin{minipage}{0.496\textwidth}
\centering
\includegraphics[width=\textwidth]{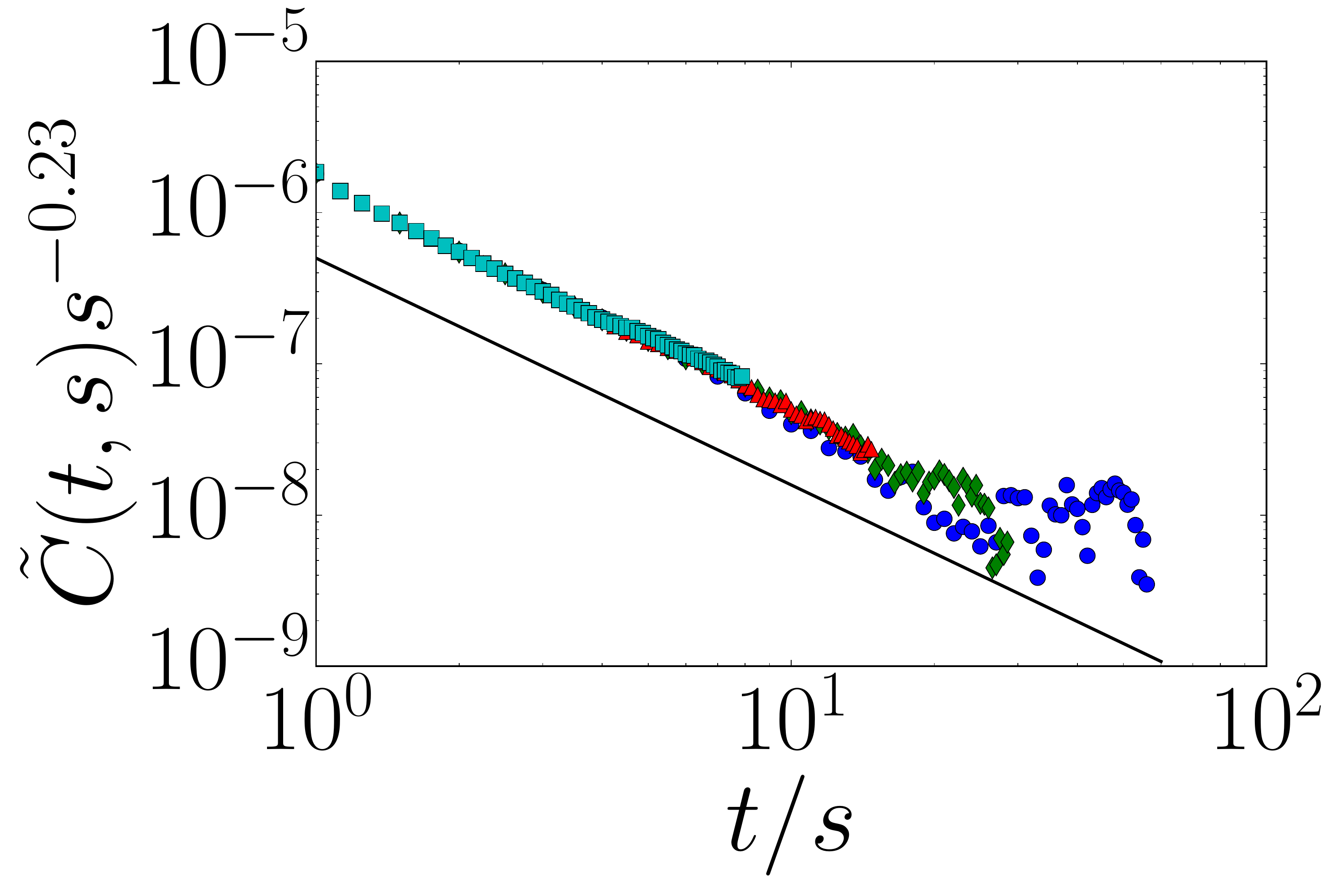}
\end{minipage}
\begin{minipage}{0.496\textwidth}
\centering
\includegraphics[width=\textwidth]{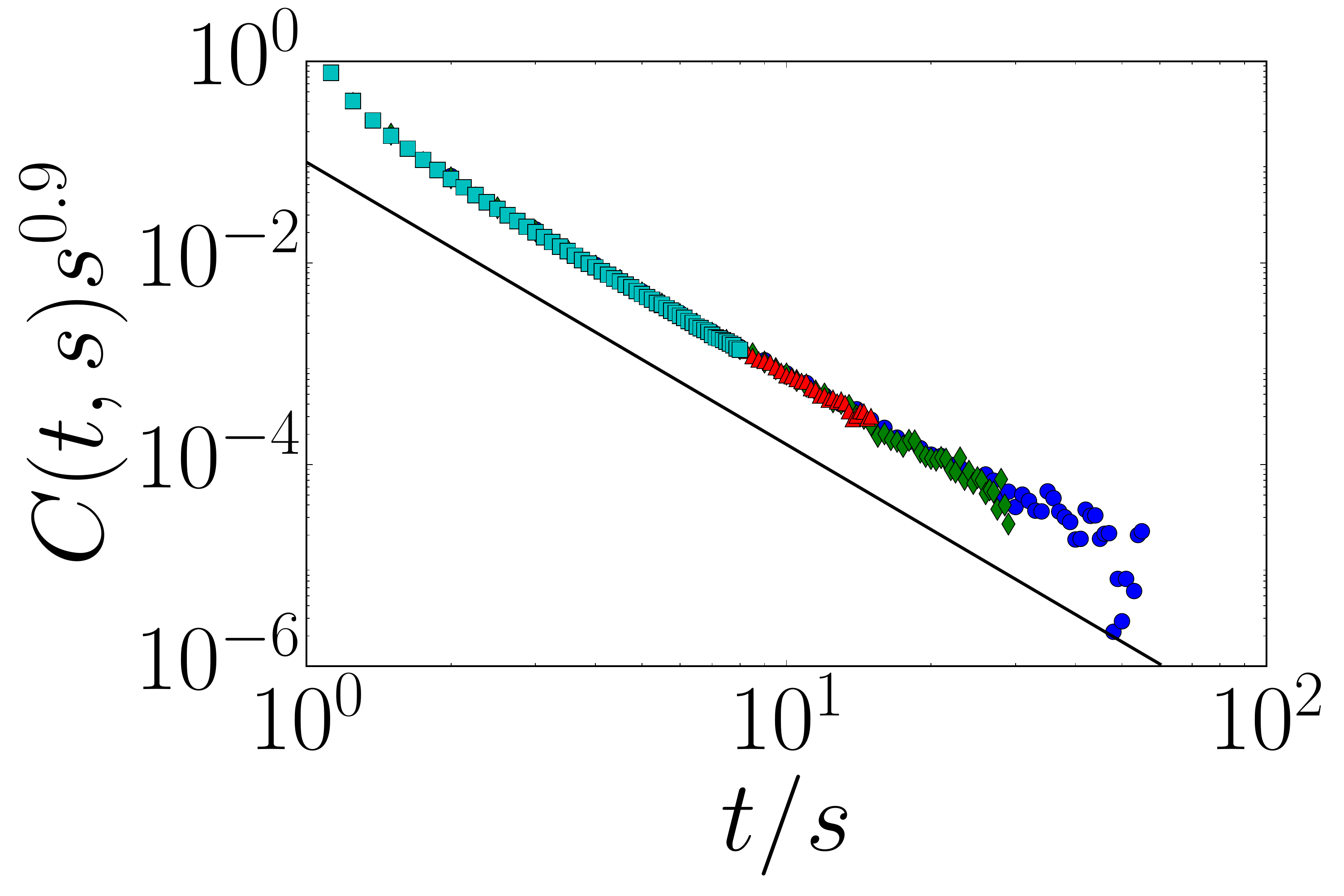}
\end{minipage}
\begin{minipage}{0.496\textwidth}
\centering
\includegraphics[width=\textwidth]{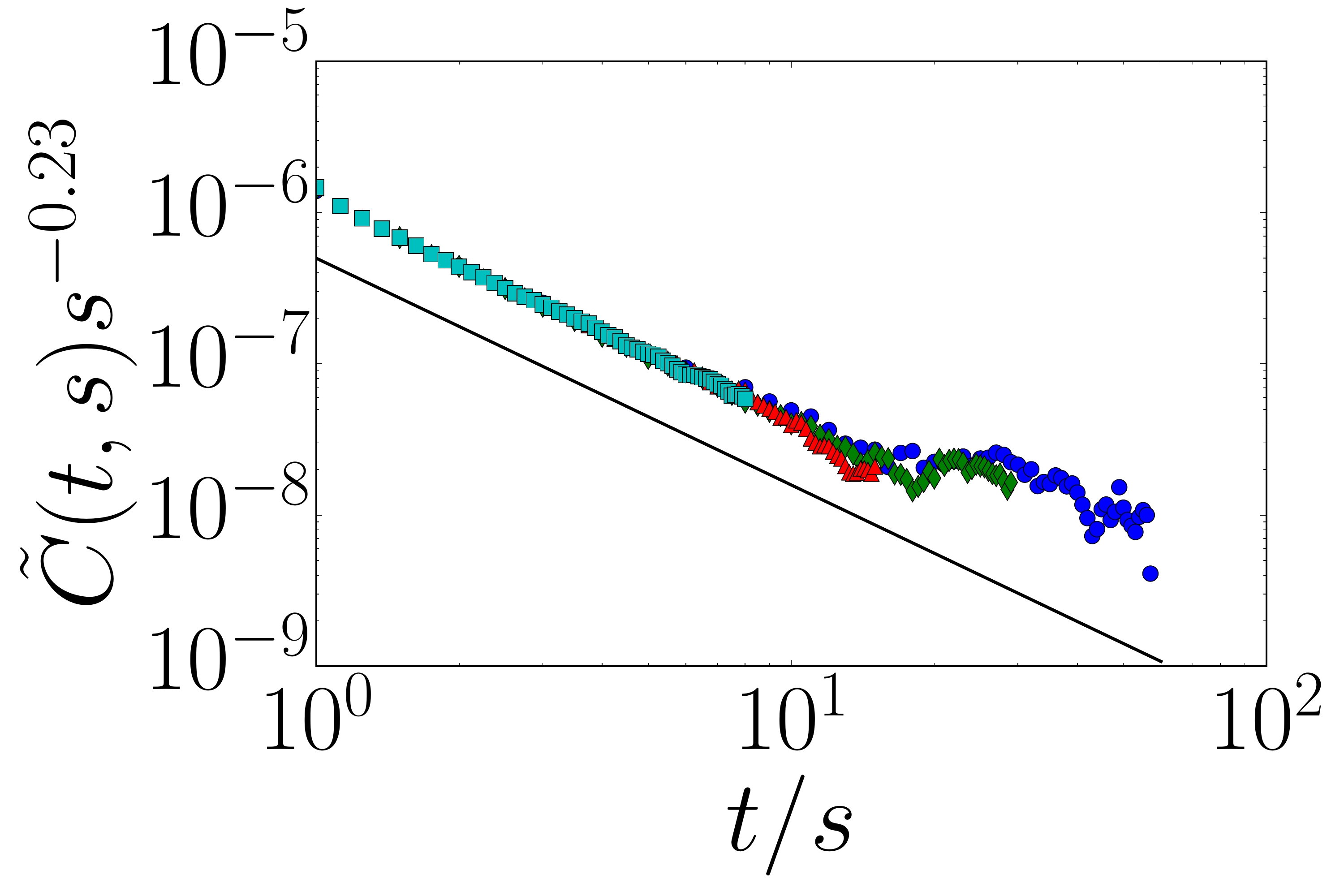}
\end{minipage}
\begin{minipage}{0.496\textwidth}
\centering
\includegraphics[width=\textwidth]{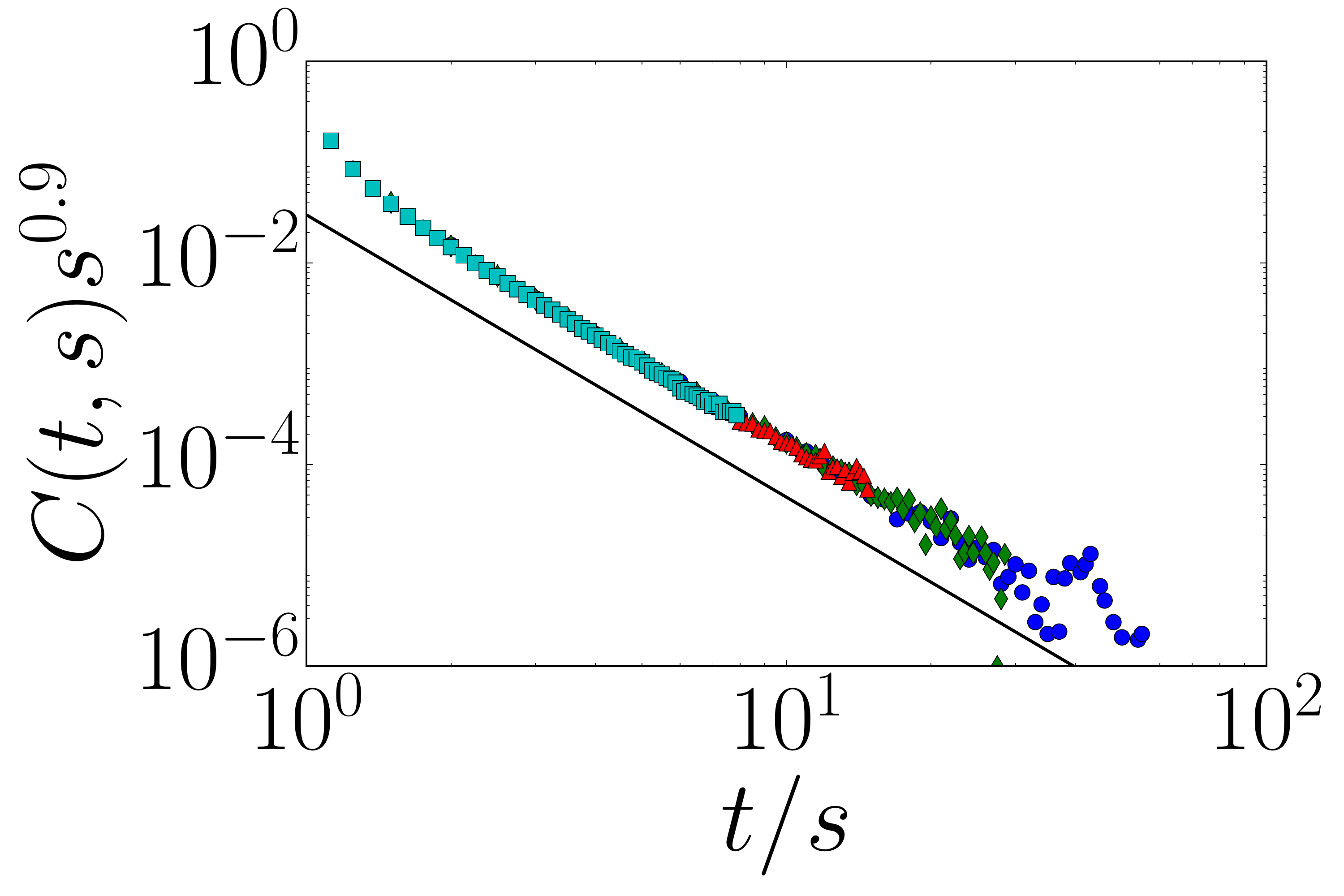}
\end{minipage}
\begin{minipage}{0.496\textwidth}
\centering
\includegraphics[width=\textwidth]{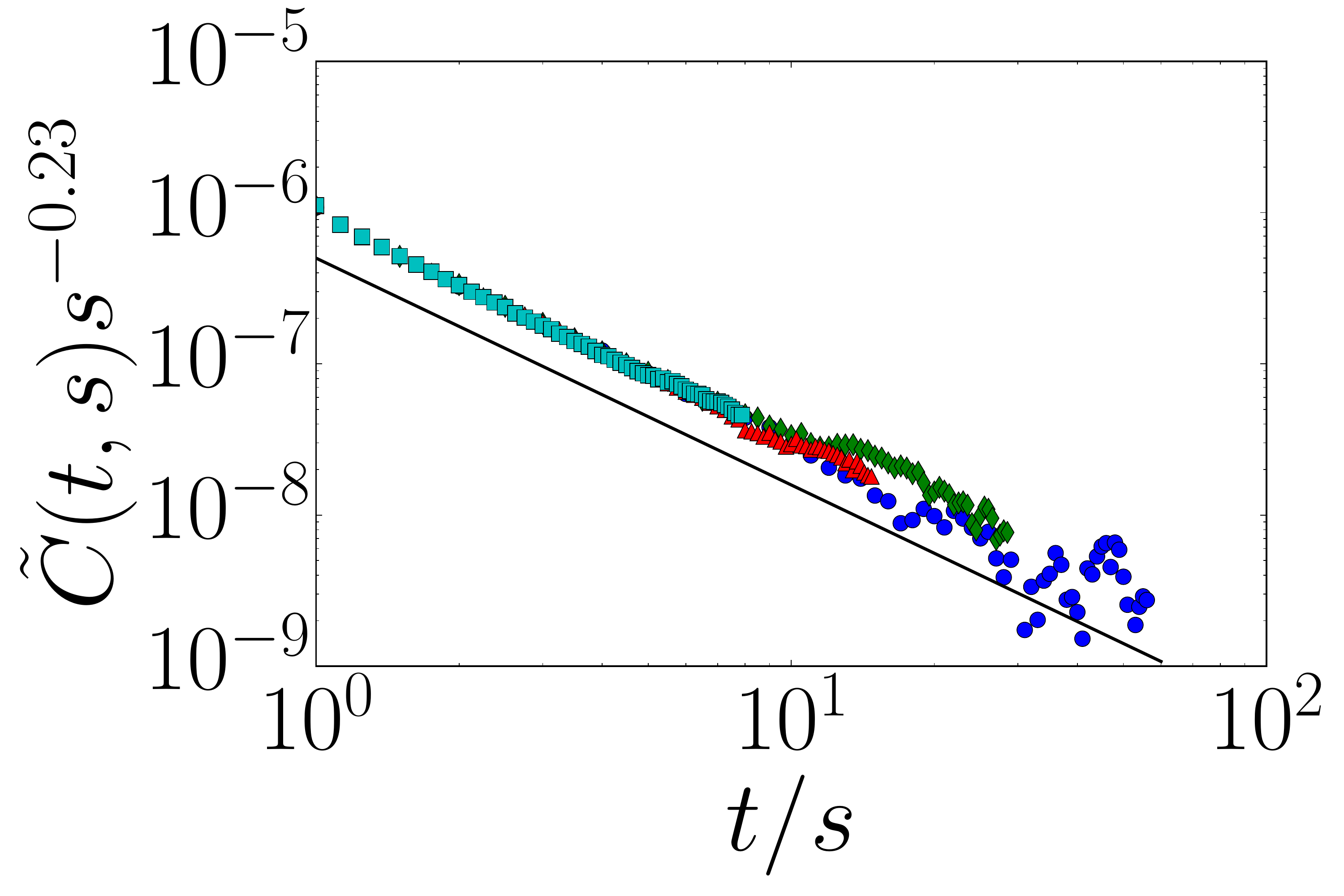}
\end{minipage}
\caption{\textbf{Connected local and global correlators at the critical point.} From top to bottom we show the data for CP$_{\lambda}$, 
CP$_{p}$ and CP$_r$. 
Left panels: Local connected correlators $C$ for different waiting times $s$. The black solid lines are guides to the eye with slopes $-2.8$. 
Right panels: Global connected correlators $\wit{C}$ for different waiting times $s$. The black solid lines are guides to the eye with slopes $-1.5$.
All data were averaged over more than $10^5$ samples.
Waiting times: blue circles $s=25$, green diamonds $s=50$, red triangles $s=100$, cyan squares $s=200$.} 
\label{fig:connected_global_local}
\end{figure}

Next, in fig.~\ref{fig:connected_global_local} we illustrate the scaling of the connected correlators $C$ and $\wit{C}$. 
The upper panel shows the local autocorrelators and their data collapse to produce the scaling form $C(t,s)=s^{-b} f_C\left(\frac{t}{s}\right)$,
as expected from eq.~(\ref{eq:c_local_scaling}). A clear scaling behaviour is seen for all three
models CP$_{\lambda}$, CP$_p$ and CP$_r$. The extracted values of the exponents agree with eqs.~(\ref{gl:scal1a}) and (\ref{gl:scal2}). 
Similarly, the lower panels display the global connected autocorrelator $\wit{C}$. The anticipated scaling from (\ref{eq:c_global_scaling}) is found
to be satisfied in all three models and the exponents agree with the expected scaling relations eqs.~(\ref{gl:scal1a}) and (\ref{gl:scal2}).
We also see that for larger values of $t/s$, the numerical values of notably the global correlators become very small and thus sensitive to effects of
purely stochastic noise in the data. 

In particular, the numerical estimates of the exponent $\wit{b}\simeq -0.23$ 
agree with the scaling relation $\wit{b}=b-d/z=2\delta-d/z=-0.231(2)$ derived in section \ref{sec:deriv_correlators}. 
Furthermore, the measured values of the exponents $\lambda_C/z$ and $\lambda_{\wit{C}}/z$ 
do reproduce the expectation from the scaling relations (\ref{gl:scal2}).

\subsection{Response function}
%
%
%
\begin{figure}
\begin{minipage}{0.496\textwidth}
\centering
\includegraphics[width=\textwidth]{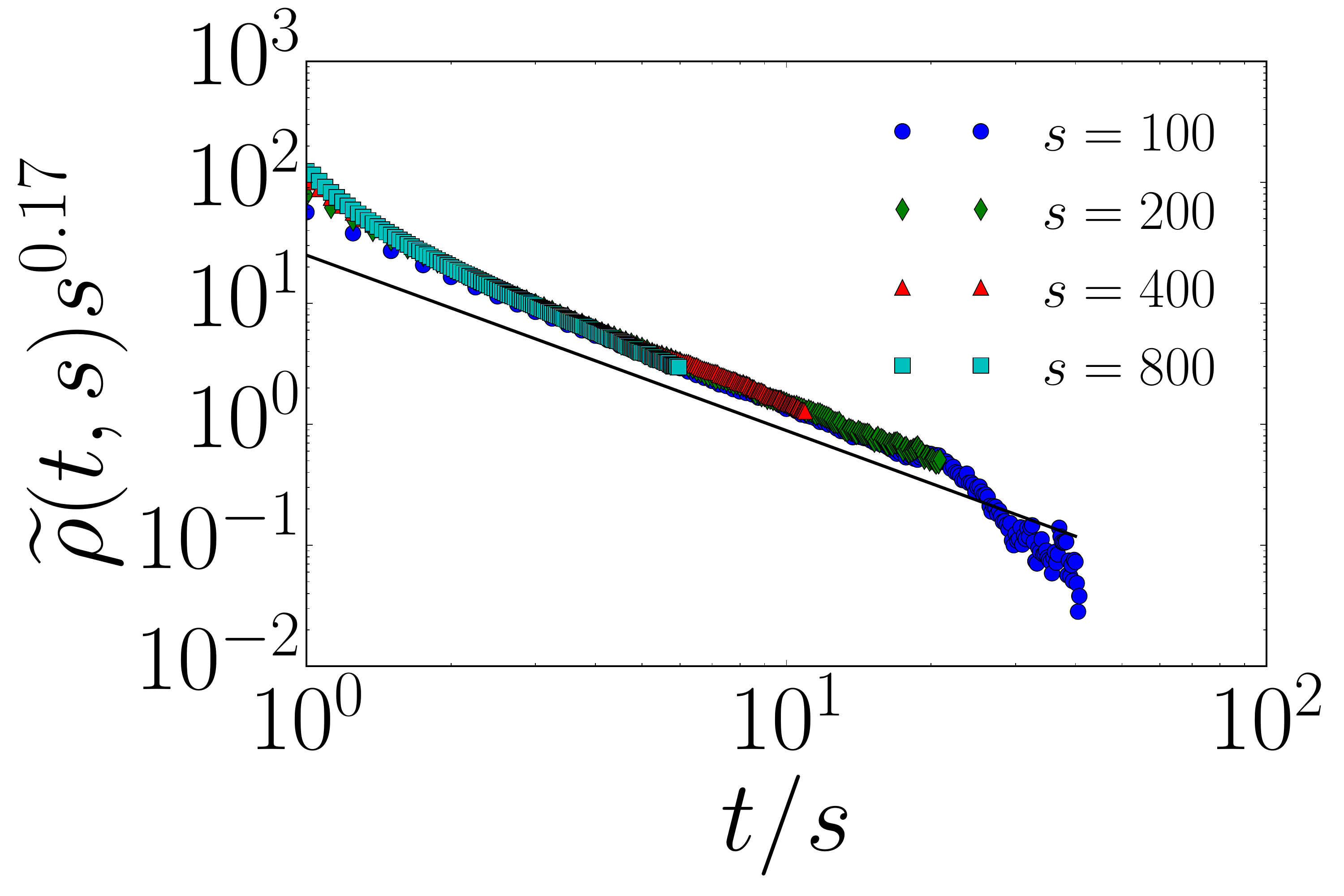}
\end{minipage}
\begin{minipage}{0.496\textwidth}
\centering
\includegraphics[width=\textwidth]{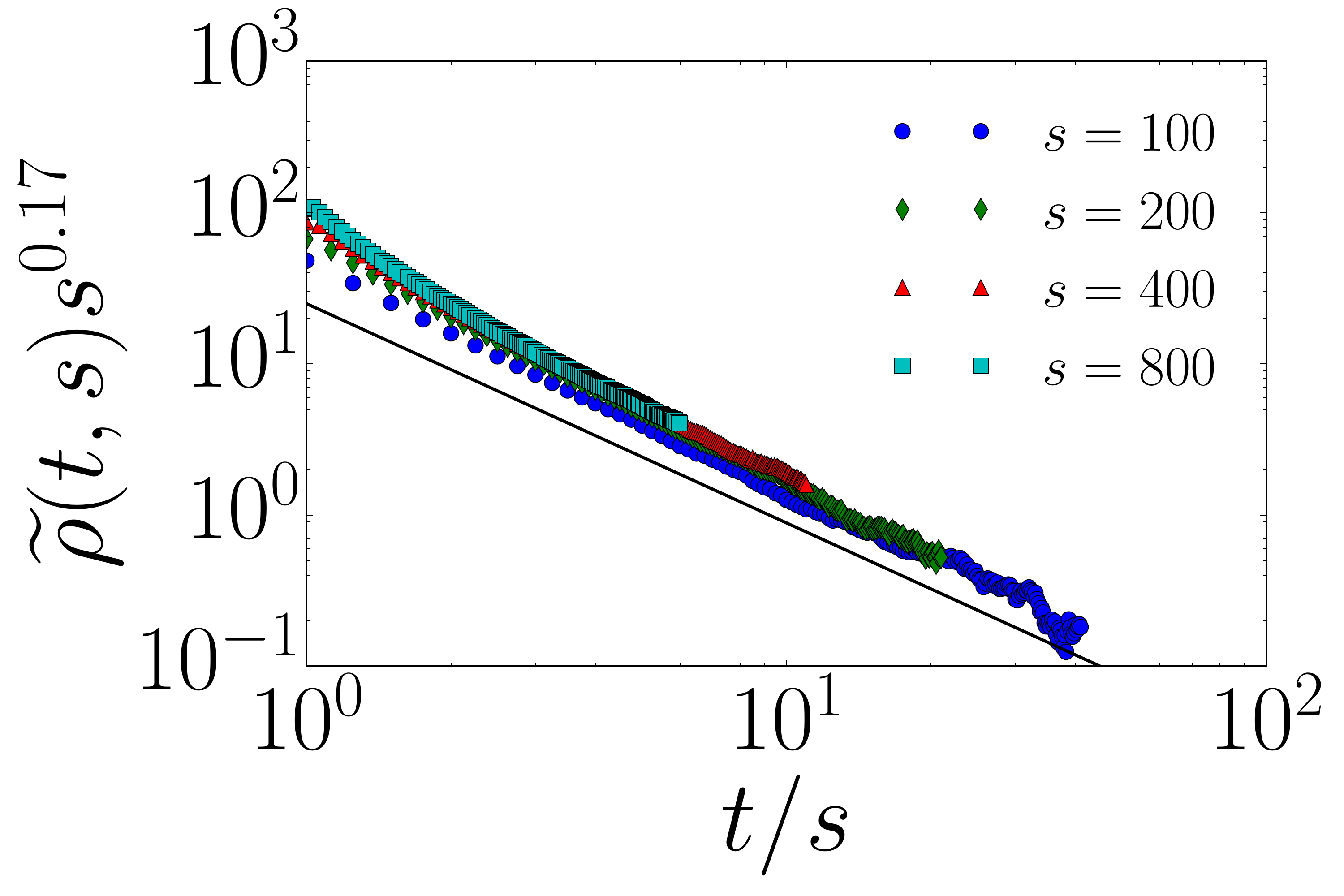}
\end{minipage}
\begin{minipage}{0.496\textwidth}
\centering
\includegraphics[width=\textwidth]{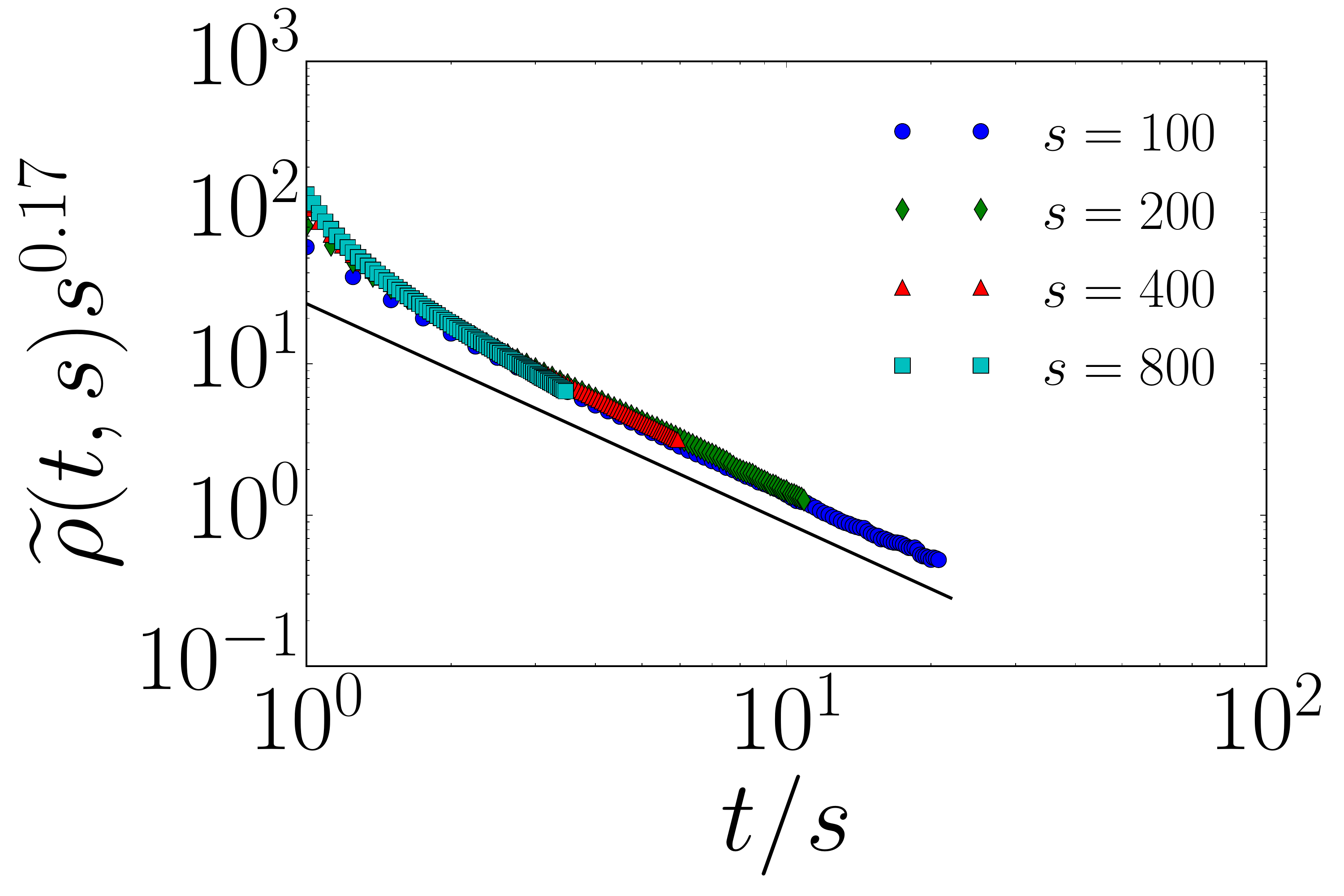}
\end{minipage}
\caption{\textbf{Integrated response functions.} 
Upper left panel: The integrated response of CP$_{\lambda}$. 
Upper right panel: The integrated response of CP$_{p}$. 
Lower panel: The integrated response of CP$_r$, for 
different waiting times $s$ respectively. All data were averaged over more than $2\cdot 10^4$ samples.
The TIR protocol has been used with $\tau_c=25$ and $h=10^{-3}$. 
The black solid lines are guides to the eye with slopes $-1.45$.} 
\label{fig:response_functions}
\end{figure}
%
%
%
In fig.~\ref{fig:response_functions} we illustrate the two-time scaling of the time-integrated global response $\wit{\rho}(t,s)$ 
for the three variants of the contact process and different waiting times $s$. 
The scaling ansatz $\wit{\rho}(t,s)=s^{-1-\wit{a}} f_{\wit{\rho}}(t/s)$ leads indeed to data collapses with $1+\wit{a}=1+a-d/z=0.17(10)$, in agreement with
the earlier findings in ref.~\cite{ramasco2004}, but not with the scaling relation (\ref{gl:scal3a}), expected from field-theory \cite{baumann2007}.
On the other hand, from the asymptotic behaviour $f_{\wit{\rho}}(y)\sim y^{-\lambda_{\wit{R}}/z}$ 
one can extract the exponent value $\lambda_{\wit{R}}/z=(\lambda_R-d)/z=1.45(5)$ 
in good agreement with both the field-theoretic scaling relation (\ref{gl:scal3b}) \cite{baumann2007}, 
as well as with earlier simulations \cite{ramasco2004}
and therefore also confirms the scaling relation \eqref{gl:scal2}.

%
%
%
\begin{figure}
\begin{minipage}{0.496\textwidth}
\centering
\includegraphics[width=\textwidth]{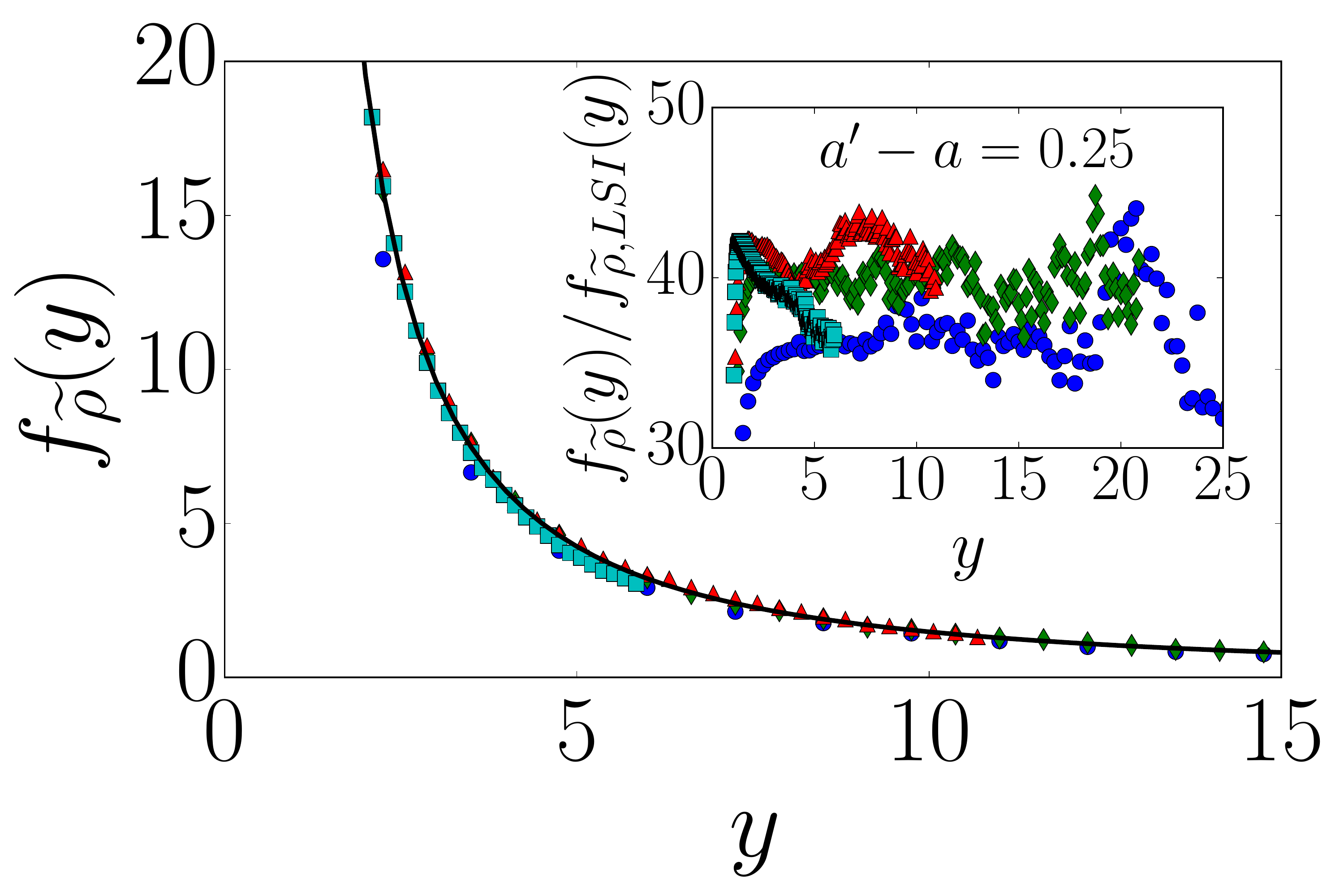}
\end{minipage}
\begin{minipage}{0.496\textwidth}
\centering
\includegraphics[width=\textwidth]{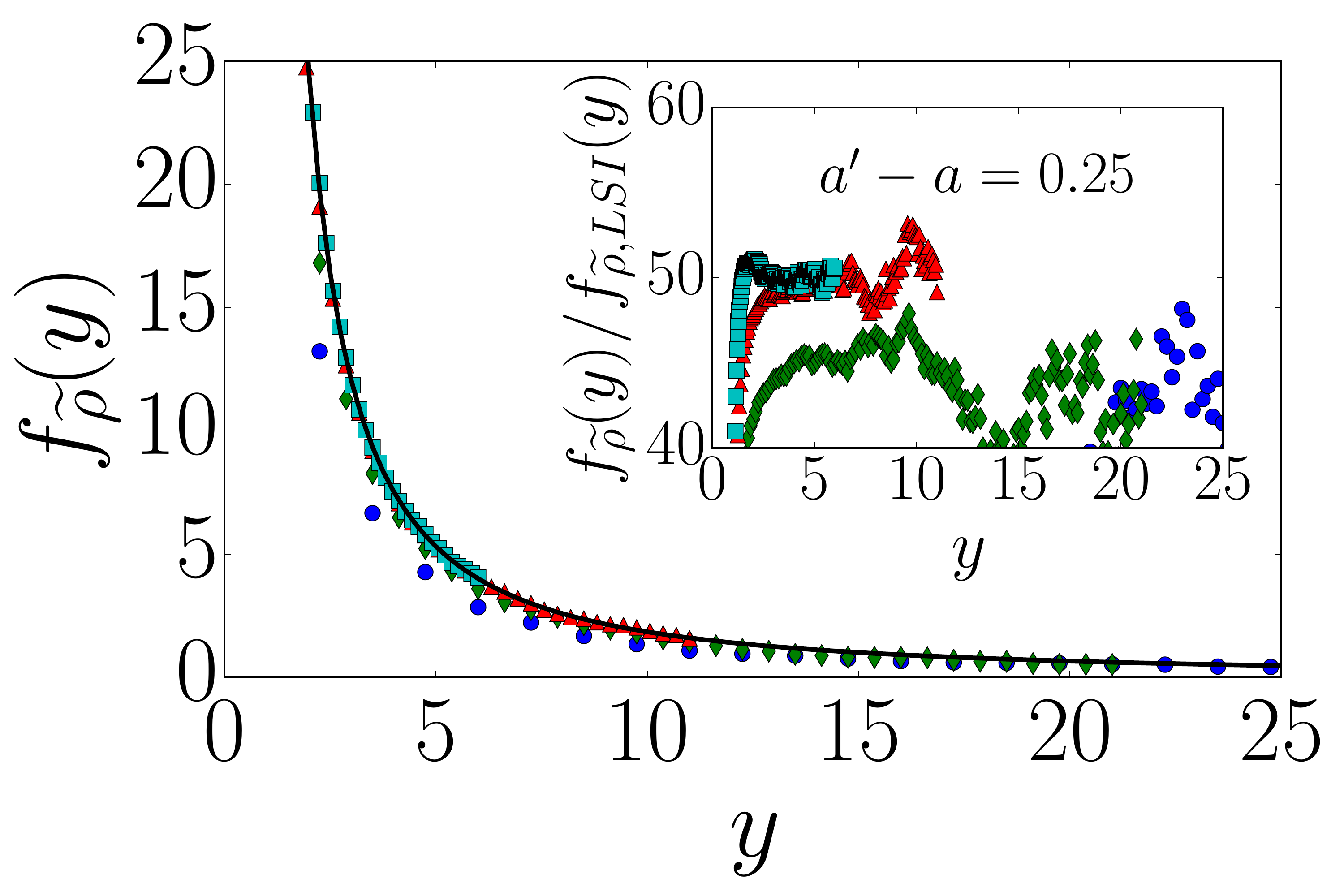}
\end{minipage}
\begin{minipage}{0.496\textwidth}
\centering
\includegraphics[width=\textwidth]{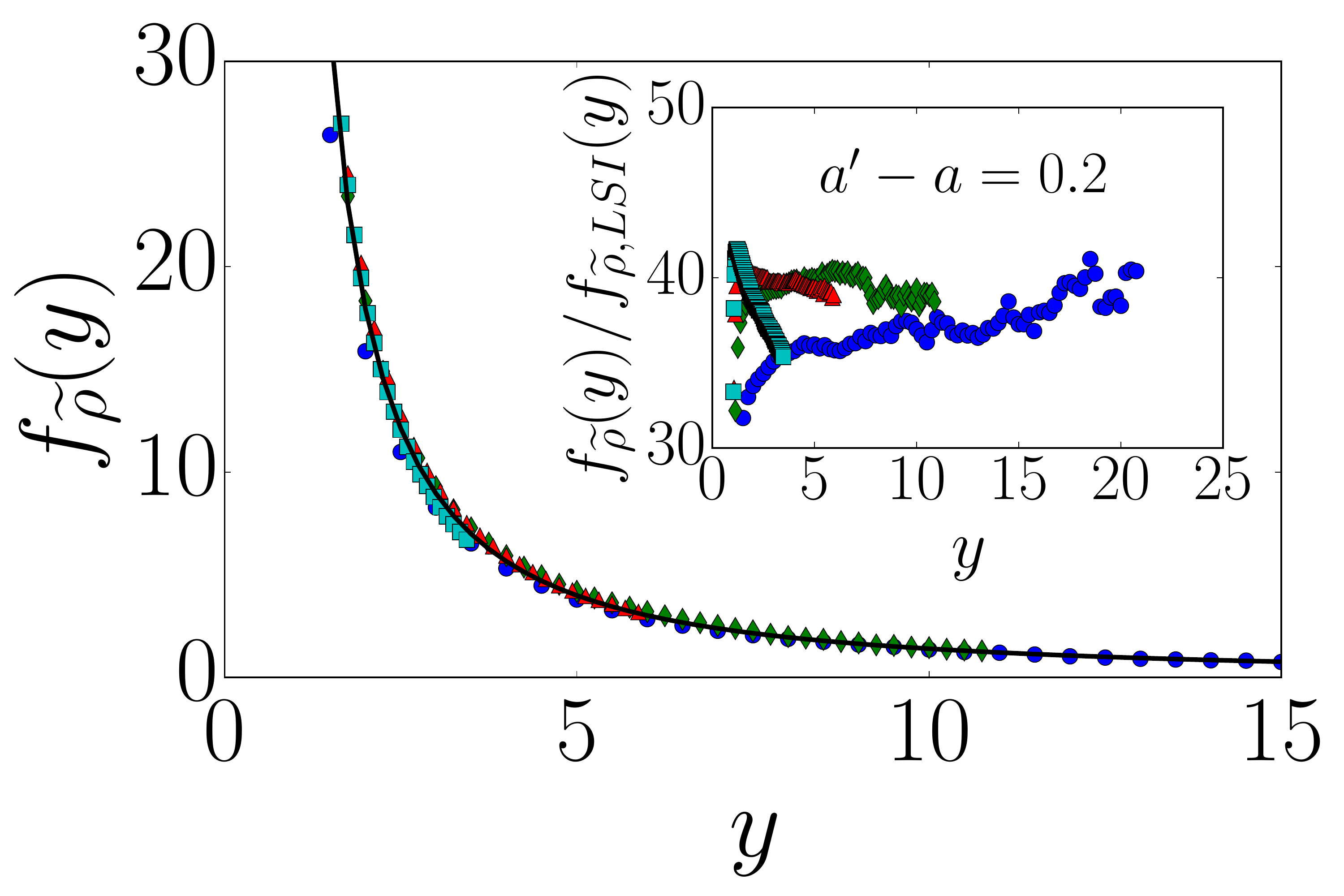}
\end{minipage}
\caption{\textbf{Scaling function of the global response.}{ Upper left panel: The scaling function $f_{\wit{\rho}}(t/s)$ of CP$_{\lambda}$. 
Upper right panel: The scaling function $f_{\wit{\rho}}(t/s)$ of CP$_{p}$. Lower panel: The scaling function $f_{\wit{\rho}}(t/s)$ of CP$_r$, for 
different waiting times $s$ respectively. The black solid lines represent the prediction (\ref{gl:lsiR}) of {\sc lsi}. 
The insets display the quotients 
$f_{\wit{\rho}}(y)/f_{\wit{\rho},{\sc LSI}}(y) = f_{\wit{\rho}}(y)\left/\left[ y^{1+a'-\lambda_R/z}(y-1)^{d/z-1-a'}\right]\right.$, 
for the values of $a'-a$ quoted in tab.~\ref{tab:cp_exp}. Waiting times: 
blue circles $s=100$, green diamonds $s=200$, red triangles $s=400$, cyan squares $s=800$.
} 
\label{fig:cp_response_function}}
\end{figure}
%
%
%

It is surprising that the estimate $a=0.3$ is quite different from the field-theoretic expectation $a=-0.1$, which means that the
scaling relation $\wit{b}=1+\wit{a}$ expected from field-theory \cite{baumann2007} is not confirmed by our numerical data for $2D$,
although the same numerical methods lead to data in agreement with that expectation for $1D$ dimensions \cite{ramasco2004,enss2004}.  
In order to test this further, we now consider the shape of the associated scaling function. This should allow to check against
a simple oversight in the measurement of the global response $\wit{\rho}$ via eq.~(\ref{eq:rho_global}) 
and its interpretation through dynamical scaling. 

Predictions on the shape of scaling functions beyond the context of a specific model can be obtained from generalisations of dynamical scaling. 
Indeed, in the context of the ageing behaviour of simple non-equilibrium magnets it has been proposed that the underlying dynamical scaling
can be extended to time-dependent conformal transformations, called {\em local scale-invariance} ({\sc lsi}) \cite{Henkel02,henkel_ageing}. 
The autoresponse takes the explicit scaling form $\wit{\rho}(t,s)\sim \wit{R}(t,s)= s^{-1-\wit{a}}f_{\wit{\rho}}(t/s)$, 
with the scaling function, for sufficiently large values of the scaling variable $y=t/s$, reads \cite{Henkel06} 
(see sect.~\ref{sec:deriv_responses} for an outline of the derivation)
\begin{equation} \label{gl:lsiR}
f_{\wit{\rho},{\sc LSI}}(y) =  f_{R,\infty}\, y^{1+a'-{\lambda_R}/{z}} (y-1)^{{d}/{z}-1-a'}.
\end{equation}
Herein, the ageing exponent $a=\wit{a}+d/z$ and the new exponent $a'$ must be chosen for the optimal description of numerical data and $f_{R,\infty}$ is a
normalisation constant.\footnote{In mean-field descriptions of the ageing of simple magnets quenched to the critical temperature $T= T_c$, 
or generically for quenches to $T<T_c$ in simple magnets, the scaling form (\ref{gl:lsiR}) holds for the two-time autoresponse with $a=a'$ \cite{henkel_ageing}. 
The exact solution of the $1D$ Glauber-Ising model, quenched to $T=0$ from a disordered initial state, produces $a=0$, $a'=-\demi$ \cite{Henkel06}.}
As reviewed in ref.~\cite{henkel_ageing} and also verified for the $1D$ directed percolation universality class, the
scaling function $f_{\wit{\rho}}(y)$ to be read from eq.~(\ref{gl:lsiR}) describes very well both simulational data \cite{enss2004,Henkel13a} 
as well as the one-loop expression derived from field-theory \cite{baumann2007}, provided the argument $y\gtrsim 2-3$. We do not wish to discuss
here whether the simple expression (\ref{gl:lsiR}) can or cannot be considered as an exact representation of the autoresponse scaling function
but rather shall use it as a simple phenomenological tool to obtain in a different way an estimate for the ageing exponent $a$, in the region
of its approximate validity 
\footnote{For $y\approx 1$, numerical data of $1D$ directed percolation require an extension to {\em logarithmic} {\sc lsi} \cite{Henkel13a,Henkel13b}.}.

In fig.~\ref{fig:cp_response_function} we show a comparison of our numerical data with the {\sc lsi}-prediction (\ref{gl:lsiR}). Indeed, for all three models
the value $a'$ needed from eq.~(\ref{gl:lsiR}) to describe the data is even larger than the value $a\approx 0.3$ read off from the simple collapse of the
data. Observing that within the numerical accuracy $a'-a\approx 0.2$ for all three models gives a quantitative form of the expected universality
of the scaling function and also illustrates that the value of $a$ needed to reproduce the shape of the scaling function appears to be even larger
than the one deduced from achieving a data collapse. This comparison is further illustrated in the insets which display estimates for the ratio of
scaling functions $f_{\wit{\rho}}(y)/f_{\wit{\rho},{\sc LSI}}(y)$, for the values of $a'-a$ quoted in tab.~\ref{tab:cp_exp}. This shows to
what extent the data collapse of dynamical scaling is actually achieved for finite values of $s$ and the available statistics. 
The value $a'-a\approx 0.2$ is comparable to the estimate $a^{\prime}-a \approx 0.27$ obtained (for non-logarithmic {\sc lsi}) 
in the $1D$ contact process \cite{Henkel06}. 
A more precise discussion
of the shape of the scaling function near $y\approx 1$ does require high-quality data with large statistics 
and for very large waiting times $s$ which at present are not available for the $2D$ contact process. 

While these considerations give additional evidence for a value of the exponent $a$ much larger than expected from the scaling
relation (\ref{gl:scal3a}), and in agreement with earlier results \cite{ramasco2004}, 
it remains an open problem why the computational technique based on eq.~(\ref{eq:int_response_def}) to find $\wit{\rho}$  
gives results in full agreement with field-theory 
in one spatial dimension and yet consistently produces a different result in two spatial dimensions.
%
%
%
\section{Derivation of scaling forms}
\label{sec:derivations}
%
%
%
We present here the scaling arguments and establish the scaling relations as defined by eqs.~(\ref{gl:scal1a}--\ref{gl:scal3}).
\subsection{Correlators}
\label{sec:deriv_correlators}
We begin with the unconnected correlators. In section~\ref{sec:correlators}, the local autocorrelator $\Gamma(t,s)$ was already treated.
Analogously, the global autocorrelator can be written as
\BEQ
\wit{\Gamma}(t,s) = \langle n(t)\rangle \langle n(s) \rangle + \wit{C}(t,s).
\EEQ
Comparing the data in figs.~\ref{fig:unconnected_global_local} and \ref{fig:connected_global_local}, we see that the numerical values of the
connected autocorrelator $\wit{C}(t,s)$ are much smaller than the values obtained for the unconnected autocorrelator, at least for the waiting times
under consideration. Therefore, we arrive at the long-time scaling $\widetilde{\Gamma}(t,s) = s^{-2\delta} n_{\infty}^2 \left( t/s\right)^{-\delta}$, in agreement
with fig.~\ref{fig:unconnected_global_local}. 

The smallness of $\wit{C}(t,s)$ can be understood from a specific symmetry of the directed percolation universality class, which is known
as {\em rapidity-reversal invariance} ({\sc rri}). This symmetry is the analogue of the time-reversal invariance of classical spin systems which relax to
an equilibrium stationary state. For directed percolation, {\sc rri} holds when the contribution of the initial state is suppressed in the
field-theory action \cite{baumann2007} and $\wit{C}(t,s)\ne 0$ only holds for an initial state distinct from the steady-state. 
On the other hand, since the long-time dynamics of systems at the critical point of their steady-state is
independent of their initial state, in critical directed percolation the influence of the initial state should disappear at long times such that
$\wit{C}(t,s)\to 0$ very rapidly. 

The discussion of the scaling of the global autocorrelator $\wit{C}(t,s)$ 
is based on the scaling assumption of the time-space correlator $C(t,s;\vec{r})$, as follows
\begin{equation}
\begin{split}
\wit{C}(t,s) 
& = \frac{1}{N^2} \sum_{i,j} \langle \left(n_i(t)-\langle n_i(t) \rangle\right) \left(n_j(s)-\langle n_j(s)\rangle\right) \rangle \\
& = \frac{1}{N} \sum_{i,j} \langle \left(n_i(t)-\langle n_i(t) \rangle\right) \left(n_0(s)-\langle n_0(s)\rangle\right) \rangle \\
& \simeq \int_{\mathbb{R}^d} \!\D\mathbf{r}\: C(t,s;\mathbf{r}) 
  = s^{-b} \int_{\mathbb{R}^d}\!\D\mathbf{r}\: F_C\left(\frac{t}{s};\frac{|\mathbf{r}|}{s^{1/z}}\right)\\
& = s^{d/z-b} \int_{\mathbb{R}^d}\!\D \mathbf{u}\: F_C\left(\frac{t}{s};|u|\right).
\end{split}
\label{eq:scaling_global}
\end{equation}
where first spatial translation-invariance and then the scaling of the time-space correlator $C(t,s;\vec{r})$ was used. 
One thus finds $\wit{b}=b-d/z=-0.231(2)$. 

The exponent $\lambda_C/z$ has also been derived using a field-theoretical approach and found to fulfil the equality 
$\lambda_C/z=1+\delta+\frac{d}{z}$ \cite{baumann2007}.
\subsection{Responses}
\label{sec:deriv_responses}
Our discussion of the scaling of the response borrows from local scale-invariance \cite{Henkel02,henkel_ageing,henkel2017}. 
A first consequence is that the full time-space scaling form factorises 
$R(t,s;\vec{r}) = R(t,s) \Phi\left( |\vec{r}| (t-s)^{-1/z}\right)$, namely into the autoresponse $R(t,s)$ 
and a universal scaling function $\Phi(u)$, which contains the space-dependence. With the normalisation $\Phi(0)=1$, we obtain
\begin{subequations}
\BEQ \label{eq:response_scaling}
\wit{\rho}(t,s) = \int_{s-\tau_c}^s \!\D u\: R(t,u) (t-u)^{d/z} \int_{\mathbb{R}^d} \!\D\vec{q}\: \Phi(|\vec{q}|) 
\sim \tau_c\, s^{d/z-1-a} \left( \frac{t}{s}\right)^{-(\lambda_R-d)/z}
\EEQ
where we used the scaling form as defined by eq.~\eqref{eq:r_scaling} of the local autoresponse 
and only retained the leading asymptotics for $y=t/s\gg 1$. The temporal integral was estimated, 
for $s\gg \tau_c$, from the mean-value theorem of integral calculus. Hence 
\begin{equation} \label{eq:response_scaling2}
\wit{\rho}(t,s)\sim s^{-(1+a-d/z)} f_{\wit{\rho}}(t/s) \;\; , \;\; f_{\wit{\rho}}(y)\sim y^{-(\lambda_R-d)/z}.
\end{equation}
\end{subequations}
This scaling form of the global integrated response actually agrees with the one proposed in ref.~\cite{baumann2007}. It also agrees with
the scaling relations of eqs.~\eqref{gl:scal2} and \eqref{gl:scal3b}. 

The shape of the scaling function $f_{\wit{\rho}}(y)$ is understood as follows. For the autoresponses, it is enough to concentrate on the
{\sc lsi}-transformation of the times. Indeed, under a change $t = \beta(t')$ of the time coordinate, a quasi-primary scaling operator $\vph(t)$ of
{\sc lsi} transforms as follows \cite{Henkel06}
\BEQ \label{gl:lsiTrans}
\vph(t) = \dot{\beta}(t')^{-x/z} \left( t'\frac{\D \ln \beta(t')}{\D t'}  \right)^{-2\xi/z}  \vph'(t')
\EEQ
where $\beta(t)$ is an arbitrary, non-decreasing, differentiable function such that also $\beta(0)=0$ \cite{Henkel06,henkel2017}. According to
Janssen-de Dominicis theory \cite{Taeuber14}, a response function $R(t,s)=\langle \vph(t) \wit{\vph}(s)\rangle$ can be formally expressed as a correlator of the
scaling operator $\vph$ associated to the order parameter (the particle-density for directed percolation) and the associated response scaling
operator $\wit{\vph}$. Because of eq.~(\ref{gl:lsiTrans}), each scaling operator is characterised by {\em two} independent scaling dimensions, 
which are denoted here by $x=x_{\vph}$ and $\xi=\xi_{\vph}$, or $\wit{x}=x_{\wit{\vph}}$, $\wit{\xi}=\xi_{\wit{\vph}}$. If time-translations were admitted 
(which would mean $\beta(t)=\beta_0\ne 0$ but this is excluded since $\beta(0)=0$) one would have $\xi=0$ 
but ageing requires that time-translation-invariance should generically be broken. The requirement of covariance of the autoresponse function
$R(t,s)$ under these transformations and (\ref{eq:response_scaling}) readily produces eq.~(\ref{gl:lsiR}), 
with $a'-a=\frac{2}{z}\left(\xi+\wit{\xi}\,\right)$ \cite{Henkel06,henkel2017}. 

%
%
%
\section{Discussion}
\label{sec:discussion}
%
%
%
\begin{table}[]
\centering
\caption[tab 2]{Contact process dynamical critical exponents for $2D$. The values of $z$ and $\delta$ are taken from ref.~\cite[tab. 4.3]{henkel08}, 
and were used to compute the values of the non-equilibrium exponents
from the scaling relations (\ref{gl:scal1a},\ref{gl:scal1b}), and (\ref{gl:scal2},\ref{gl:scal3}) \cite{baumann2007}, labelled `scaling'. 
The experimental data `exp' are from the turbulent liquid crystal MBBA \cite{Takeuchi09}. 
The numbers in brackets indicate the estimated uncertainty in the last given digit(s).
For comparison, results of the earlier simulation \cite{ramasco2004} are also listed.}
\label{tab:cp_exp}
\begin{tabular}{|l|lll|l|l|} \hline
Exponent    & \multicolumn{1}{c}{CP$_{\lambda}$} & \multicolumn{1}{c}{CP$_{p}$}   & \multicolumn{1}{c|}{CP$_r$} & \multicolumn{1}{c|}{exp} 
& \multicolumn{1}{c|}{scaling}\\ \hline\hline
~$z$                          &~~$1.7660(16)$     &~~$1.7660(16)$                  &~~$1.7660(16)$         & ~~$1.74(8)$ & ~~$1.7660(16)$~ \\ 
~$\delta$                     & ~~~$0.4505(10)$   & ~~~$0.4505(10)$                & ~~~$0.4505(10)$       & ~~$0.46(5)$ & ~~$0.4505(10)$ \\ \hline
~$\lambda_\Gamma/z$           & ~~~$0.4505(10)$   & ~~~$0.4505(10)$                & ~~~$0.4505(10)$       &             & ~~$0.4505(10)$ \\ 
~$\lambda_{\wit{\Gamma}}/z$   & ~~~$0.4505(10)$   & ~~~$0.4505(10)$                & ~~~$0.4505(10)$       &             & ~~$0.4505(10)$ \\ 
~$\lambda_{C}/z$              & ~~~$2.8(2)$       & ~~~$2.8(2)$                    & ~~~$2.8(2)$           & ~~$2.5(3)$  & ~~$2.58(2)$    \\ 
                              &                   & ~~~$2.8(3)$ \cite{ramasco2004} &                       &             &                \\ 
~$\lambda_{\wit{C}}/z$        & ~~~$1.5(1)$       & ~~~$1.5(1)$                    & ~~~$1.5(1)$           &             & ~~$1.4505(10)$ \\ 
~$b$                          & ~~~$0.901(2)$     & ~~~$0.901(2)$                  & ~~~$0.901(2)$         & ~~$0.9(1)$  & ~~$0.901(2)$   \\ 
~$\wit{b}$                    & $-0.232(2)$       & $-0.232(2)$                    & $-0.232(2)$           &             & $-0.232(2)$    \\ \hline
~$\lambda_{\wit{R}}/z$        & ~~~$1.45(5)$      & ~~~$1.50(5)$                   & ~~~$1.45(5)$          &             & ~~$1.4505(10)$ \\ 
                              &                   & ~~~$1.62(10)$ \cite{ramasco2004} &                     &             &                \\ 
~$1+\wit{a}$                  & ~~~$0.17(10)$     & ~~~$0.17(10)$                  & ~~~$0.17(10)$         &             & $-0.232(2)$    \\ 
~$a'-a$                       & ~~~$0.25(15)$     & ~~~$0.25(15)$                  & ~~~$0.20(15)$         &             & ~~$0$          \\ \hline
\end{tabular}
\end{table}
%
%
%
Characterising phase transitions and universal features of spreading models have given invaluable insights in 
different spreading dynamics \cite{grassberger83,marro05,henkel08,boettcher171}. 
In our study, we focused on a classification of the universality of dynamical properties of the contact process in two dimensions by considering three 
different numerical models. In addition to earlier studies, we also considered the scaling of global correlators. 
Our results on the local correlators and the response function are in good agreement with earlier results \cite{ramasco2004,chen16}. 
We derived an analytical relation between local and global correlators.
All scaling exponents are summarised in tab.~\ref{tab:cp_exp} and the expected universality between the three models considered in this work is clearly confirmed. 
All theoretically expected exponent scaling relations 
eqs.~(\ref{gl:scal1a}--\ref{gl:scal3}) were confirmed, with the only exception of (\ref{gl:scal3a}). A further analysis of the
scaling function of the global response function revealed that this shape is numerically very well described by a simple version of
local scale-invariance but also confirms the violation of (\ref{gl:scal3a}). 
This is surprising, since (\ref{gl:scal3a}) follows from standard field-theoretical arguments and the analogous numerical analysis in one
spatial dimension is known to produce results consistent with (\ref{gl:scal3a}).
We have shown that the scaling of the unconnected correlator is an appropriate tool to accurately determine critical points. 
Future studies might study the local integrated response to also test the scaling relation as defined by eq.~\eqref{gl:scal1a}.
Such studies might also explore the relaxation dynamics in a parameter regime where the general contagion model CP$_r$ 
allows for two coexisting stationary states in a cusp catastrophe phase space \cite{boettcher162, boettcher171}.
\section*{Acknowledgments}
We warmly thank Jos\'{e} Javier Ramasco for helpful comments and suggestions. 
We acknowledge financial support from the ETH Risk Center and ERC Advanced grant number FP7-319968 FlowCCS of the European Research Council. 
We also thank the Instituto Nacional de Ci\^{e}ncia e Tecnologia de Sistemas Complexos (INCT-SC) for financial support. 
%
%
%
%
%
%
%
%
%
%

%
%
%
\end{document}